\documentclass{ws-ijgmmp}

\usepackage{xcolor}
\definecolor{revisionwarm}{RGB}{166,88,56}

\usepackage[sort,compress]{cite}
\usepackage{bm}
\usepackage{booktabs}
\usepackage{placeins}
\usepackage[verbose,hypertexnames=false]{hyperref}
\usepackage{orcidlink}
\hypersetup{
  colorlinks=false,
  allbordercolors=blue,
  pdfborderstyle={/S/U/W 1},
  pdftitle={A dark-matter-sensitive orientation clock near rotating black holes},
  pdfauthor={Mohsen Fathi},
  pdfsubject={Geometric relativistic orientation clock and synthetic timing forecast},
  pdfkeywords={differential geometry, rotating black holes, dark matter environments, extended-body dynamics, relativistic timing}
}

\newcommand{\dd}{\mathrm{d}}
\newcommand{\DM}{\mathrm{DM}}
\newcommand{\flip}{\mathrm{flip}}
\newcommand{\Kerr}{\mathrm{Kerr}}
\newcommand{\cNFW}{\mathrm{cored\mbox{-}NFW}}

\newcommand{\vect}[1]{\bm{#1}}

\def\wsdraftnote{}

\begin{document}

\markboth{M. Fathi}{Orientation clock near rotating black holes}
\catchline{}{}{}{}{}

\title{A dark-matter-sensitive orientation clock near rotating black holes}

\author{Mohsen Fathi\,\orcidlink{0000-0002-1602-0722}}
\address{Centro de Investigaci\'{o}n en Ciencias del Espacio y F\'{i}sica Te\'{o}rica (CICEF),\\ Universidad Central de Chile, La Serena 1710164, Chile\\
\email{mohsen.fathi@ucentral.cl}}

\maketitle

\begin{abstract}
A coherent triaxial structure has an intrinsic orientation clock because rotation about its intermediate principal axis is unstable. I ask how this clock changes when the structure moves on a circular equatorial orbit near a rotating black hole in an effective dark matter (DM) environment. The calculation uses a comoving orthonormal tetrad, the electric part of the Riemann tensor, Euler--quaternion dynamics, and the conversion from proper time to stationary coordinate time. The clock shift separates exactly into a local proper-time response and a geometric time-conversion response. I compute a direct grid of 600 DM models for Einasto, regularized cored Navarro--Frenk--White, and Hernquist profiles. Across the scanned domain, the total fractional frequency shift lies between about $-3\times10^{-7}$ and $-3\times10^{-4}$. For the fiducial body model, most of the shift comes from the time conversion. Synthetic timing tests show that a shift near $10^{-4}$ is recoverable only when the clock is highly coherent and its Kerr baseline is known independently. This gives a theoretical timing diagnostic and a synthetic observational forecast. I do not fit real data.
\end{abstract}



\section{Introduction}
\label{sec:intro}

A clock near a rotating black hole is more than a single frequency. The value assigned to it depends on the local dynamics, the observer, and the way the signal reaches a distant detector. Orbital motion, radial and vertical epicyclic motion, periastron precession, disk oscillations, resonances, and Lense--Thirring (LT) precession are well-known examples in compact-object timing \cite{Stella:1999ct,Abramowicz:2001bi,Nowak:1999,Kato:2008,Ingram:2009eb,Motta:2014qpa,IngramMotta:2020}. Their use in data analysis also requires a model for coherence and radiation. A narrow peak can be useful, but it does not by itself reveal the physical mechanism \cite{Remillard:2006fc,vanDerKlis:2006,BachettiHuppenkothen:2024}.

Here I consider a different clock. A freely rotating triaxial body with principal moments $I_1<I_2<I_3$ is unstable near rotation about the intermediate axis. This is the tennis-racket or Dzhanibekov effect. It follows from the Euler equations and has a clear geometric structure in angular-momentum space \cite{Goldstein:2002,Landau:1976,Montgomery:1991,VanDamme:2017,Mardesic:2020}. Dissipation can weaken the flips and eventually drive the body toward stable rotation \cite{deLaTorre:2024}. The instability itself is therefore classical and intrinsic. A black hole or a DM halo does not create it.

The question is more specific. Suppose that a coherent triaxial element already carries this intrinsic orientation clock. How does the clock change as the element moves through a weakly DM-dressed rotating geometry? Several results give qualitative motivation for such an element. Cooling self-gravitating disks can develop spiral density waves and may fragment when cooling is sufficiently fast \cite{Gammie:2001,Cossins:2009}. Orbiting hot spots are also used as reduced timing models, and resistive general relativistic magnetohydrodynamics (GRMHD) simulations can form macroscopic plasmoids close to black holes \cite{Schnittman:2005,Ripperda:2020}. These results do not establish the long-lived triaxial element assumed here. They only suggest that a finite non-axisymmetric structure is a reasonable reduced dynamical probe. The element is not the whole accretion disk, and I do not treat the disk as a rigid body. Its formation, survival, and radiation require a separate fluid calculation.

The dark matter environment is also uncertain. Einasto, Navarro--Frenk--White, Burkert, Hernquist, and related cored profiles are widely used to describe galactic mass distributions \cite{Einasto:1965,Navarro:1996gj,Navarro:1996he,Hernquist:1990be,Burkert:1995yz}. A black hole may steepen the inner distribution through adiabatic growth, while scattering, mergers, annihilation, and the growth history may reduce a spike \cite{Gondolo:1999ef,Ullio:2001fb,Sadeghian:2013laa}. Environmental effects around black holes have been studied through waveforms, orbital dynamics, images, and effective metrics \cite{Barausse:2014tra,Kavanagh:2020cfn,Cardoso:2021wlq,Figueiredo:2023gas}. The profiles used below are controlled effective backgrounds rather than measured profiles of a specific source.

I use a Kerr-like rotating ansatz built from a spherical mass function. Similar constructions are often used as phenomenological probes \cite{Newman:1965tw,AzregAinou:2014pra,Xu:2018wow,Xu:2020bxi,Jusufi:2019ltj}. The metric serves only as a phenomenological background. It is not presented as an exact solution of the Einstein equations produced by a specified rotating DM stress-energy tensor. It maps a density profile to an enclosed mass, a local geometry, a tidal tensor, and a clock shift.

The reduced body model is motivated by the relativistic theory of extended bodies. The Mathisson--Papapetrou--Dixon equations show how spin and multipole moments couple to curvature \cite{Mathisson:1937zz,Papapetrou:1951pa,Dixon:1970zza,Dixon:1974c,EhlersRudolph:1977}. Quadrupole effects in Kerr spacetime have been studied in several forms \cite{Bini:2009,BiniGeralico:2014,Bini:2015zya,Han:2016cdm,SteinhoffPuetzfeld:2012}. I do not solve that complete covariant problem. Instead, I project the curvature into a comoving frame and use the electric tidal tensor in a reduced body-frame torque. Its strength is controlled by the effective coefficient $K_{\rm tidal}$, which is not calibrated to a specific fluid or material model. The relative size of the two clock channels therefore depends on the chosen coupling. It is not a universal property. Section~\ref{sec:robustness} tests this dependence.

A central step is to keep two effects separate: the change of the orientation dynamics in proper time and the change in the conversion to stationary coordinate time. Their exact relation follows from the circular-orbit clock definition. For the fiducial model, the calculation shows that the time-conversion part is much larger than the local tidal part.

\begin{figure}[!t]
\centering
\includegraphics[width=\textwidth]{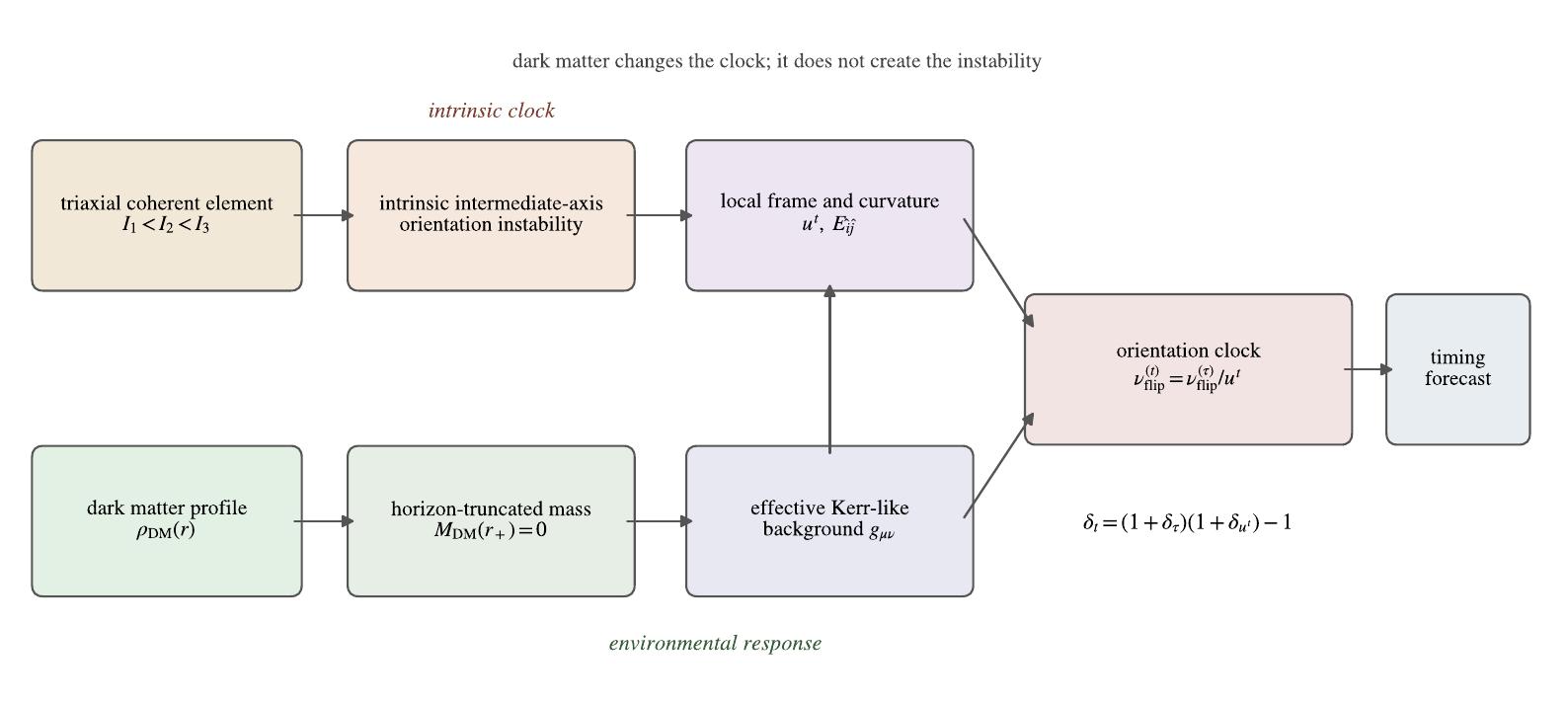}
\caption{Structure of the calculation. The upper branch is the intrinsic intermediate-axis clock of a coherent triaxial element. The lower branch is the response of a horizon-truncated DM profile. The two branches meet through the comoving frame and the local curvature. Dark matter changes the clock but does not create the instability. The last step tests the predicted shift in synthetic timing data.}
\label{fig:architecture}
\end{figure}

The second aim is a synthetic observational forecast. I convert the dimensionless clock into illustrative physical frequencies, generate orientation-modulated light curves, measure their spectral peaks, add noise and finite coherence, and test model separation. I also compare the expected shifts with published timing scales. {Figure~\ref{fig:architecture} summarizes the full path.} Together, the geometry, direct dynamics, robustness tests, and timing application form one self-contained framework.

\section{Effective Dark-Matter-Dressed Rotating Geometry}
\label{sec:background}

I use geometrized units $G=c=M=1$ unless physical units are written explicitly. Here $M$ is the black hole mass, $a$ is the spin parameter, and $\chi=a/M$.

\subsection{Kerr reference geometry and inner boundary}

For the vacuum Kerr reference geometry, the radial function is
\begin{equation}
 \Delta_{\rm K}(r)=r^2-2Mr+a^2.
\label{eq:deltak}
\end{equation}
Its outer root is
\begin{equation}
 r_+^{\rm K}=M\left(1+\sqrt{1-\chi^2}\right).
\label{eq:rplus}
\end{equation}
I use $r_+^{\rm K}$ as the fixed inner boundary of the halo mass integral. The boundary is therefore fixed by the Kerr reference geometry before the effective dark matter metric is constructed.

\subsection{Profiles and one common mass convention}

The Einasto density is
\begin{equation}
 \rho_{\rm E}(r)=\rho_e\exp\left[-\frac{2}{\alpha_E}
 \left(\left(\frac{r}{r_e}\right)^{\alpha_E}-1\right)\right],
\label{eq:einasto}
\end{equation}
with $\alpha_E=0.18$ and $r_e=80M$. The regularized cored Navarro--Frenk--White (cored-NFW) profile is
\begin{equation}
 \rho_{\cNFW}(r)=\frac{\rho_s r_s^3}{(r+r_c)(r+r_s)^2},
\label{eq:cnfw}
\end{equation}
with $r_s=80M$ and $r_c=5M$. I also use a Hernquist control profile,
\begin{equation}
 \rho_{\rm H}(r)=\frac{M_h}{2\pi}\frac{a_h}{r(r+a_h)^3},
\label{eq:hernquist}
\end{equation}
with $a_h=80M$ before normalization.

All profiles use the same enclosed-mass definition,
\begin{equation}
 M_{\DM}(r)=4\pi\int_{r_+^{\rm K}}^{r}\rho_{\DM}(x)x^2\dd x,
 \qquad M_{\DM}(r_+^{\rm K})=0.
\label{eq:mass}
\end{equation}
The normalization is
\begin{equation}
 \mu_{\DM}=\frac{M_{\DM}(200M)}{M}.
\label{eq:mu}
\end{equation}
The lower limit in Eq.~\eqref{eq:mass} is a model choice. It prevents an arbitrary continuation of a galactic profile through the horizon from being interpreted as a physical horizon displacement. The construction therefore studies the exterior environmental response and does not infer a DM-driven horizon shift from an interior mass convention.

\subsection{Kerr-like metric}

The spherical seed function is
\begin{equation}
 f(r)=1-\frac{2m(r)}{r},\qquad m(r)=M+M_{\DM}(r).
\label{eq:seed}
\end{equation}
I place this mass function in the Kerr-like line element
\begin{align}
 \dd s^2={}&-\left(1-\frac{2m(r)r}{\Sigma}\right)\dd t^2
 -\frac{4a m(r)r\sin^2\theta}{\Sigma}\dd t\dd\phi
 +\frac{\Sigma}{\Delta}\dd r^2+\Sigma\dd\theta^2 \nonumber\\
 &+\sin^2\theta\left[r^2+a^2+
 \frac{2a^2m(r)r\sin^2\theta}{\Sigma}\right]\dd\phi^2,
\label{eq:metric}
\end{align}
where
\begin{equation}
 \Sigma=r^2+a^2\cos^2\theta,\qquad
 \Delta=r^2+a^2-2m(r)r.
\label{eq:sigmadelta}
\end{equation}
The Kerr metric is recovered exactly when $M_{\DM}=0$ \cite{Kerr:1963ud,Carter:1968rr,Bardeen:1972fi,Chandrasekhar:1983}. Because $M_{\DM}(r_+^{\rm K})=0$, the effective function $\Delta$ also vanishes at $r_+^{\rm K}$. This boundary condition alone would not exclude a second zero farther out. I therefore scan $\Delta(r)$ independently for all 60 profile--spin--normalization backgrounds used by the production grid. In every case $\Delta(r)>0$ on $(r_+^{\rm K},200M]$. The smallest right derivative at the reference horizon is $\partial_r\Delta|_{r_+^{\rm K}}=0.6244$, reached for the Hernquist profile with $\chi=0.95$ and $\mu_{\DM}=0.05$. Thus $r_+^{\rm K}$ remains the outermost zero throughout the scanned domain.

Equation~\eqref{eq:metric} defines an effective stationary and axisymmetric geometry, together with its Riemann tensor, circular orbits, and local orthonormal frames. I use it only to test how a chosen radial mass function changes a relativistic clock. It does not derive a unique rotating matter source or the strong-field behavior of a specific dark matter particle model.

\subsection{Local frame and geometric meaning}

A local tidal measurement needs an observer and a frame. Fermi normal coordinates and parallel-propagated frames give the standard geometric basis for such measurements \cite{ManasseMisner:1963,Marck:1983}. In the present circular problem I construct an orthonormal tetrad whose timelike leg is the orbital four-velocity. The tensor eigenvalues do not change under a spatial rotation of this tetrad, but body-frame torque components do depend on the orientation of the triaxial element relative to the tidal eigendirections.

Proper time on the orbit is invariant. By contrast, a frequency written in the Boyer--Lindquist-like time $t$ uses the stationary time coordinate of the effective geometry. The coordinate is normalized by the large-radius limit $g_{tt}\rightarrow-1$. It is therefore useful for a distant timing comparison, but it is not a local scalar and it is not yet an observed photon arrival time. The latter would also require photon propagation and an emission model. Keeping the two time variables separate is central to the analysis.

\begin{figure}[!t]
\centering
\includegraphics[width=0.84\textwidth]{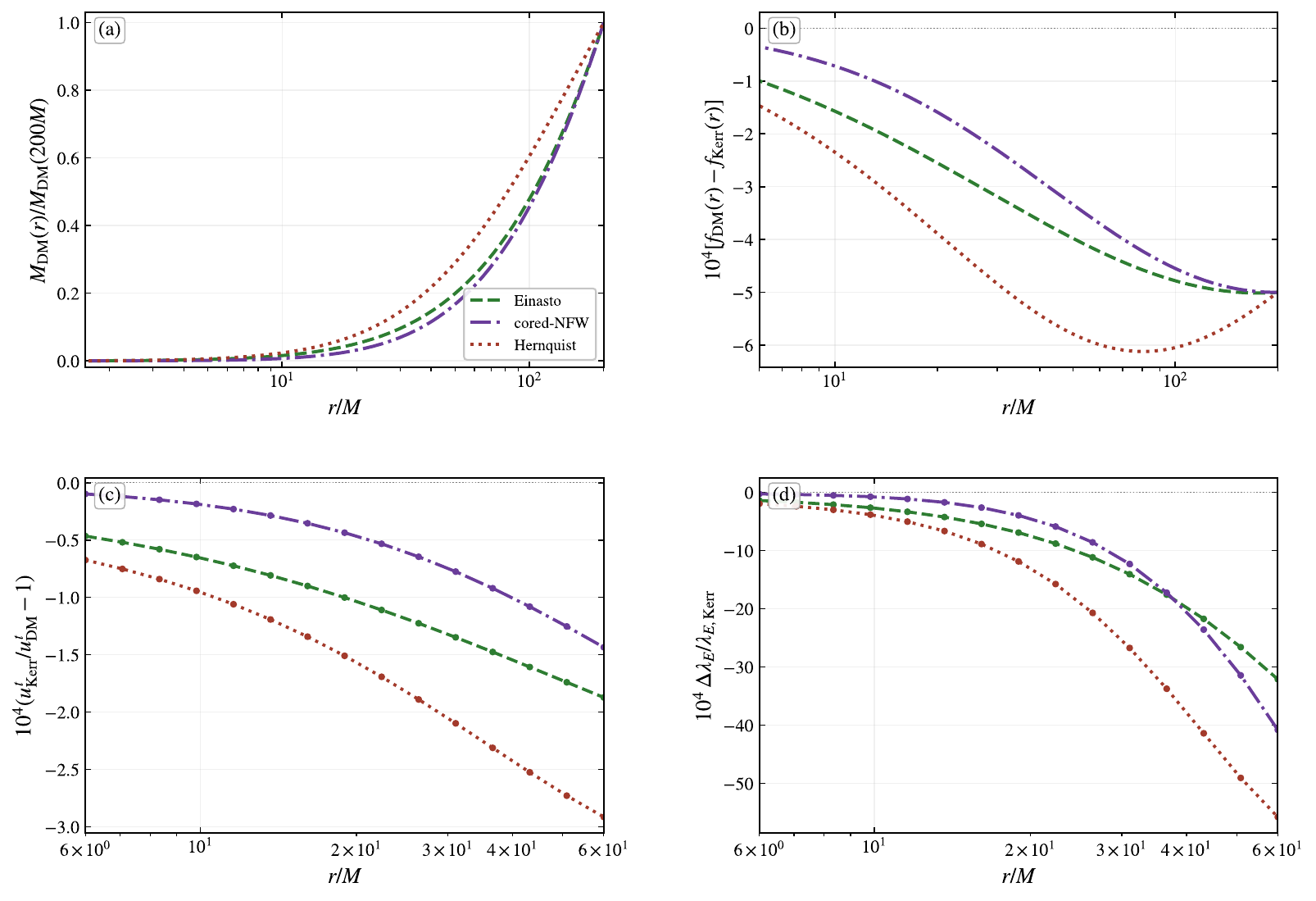}
\caption{Environmental response for $\chi=0.8$ and $\mu_{\DM}=0.05$. The panels show (a) the enclosed mass, (b) the change of the seed function relative to Kerr, (c) the time-conversion contribution, and (d) the change of the dominant electric-tidal eigenvalue. All profiles have the same mass at $200M$, but different radial mass buildup.}
\label{fig:background}
\end{figure}

Figure~\ref{fig:background} follows the response from the profile to the local geometry. The profiles have the same enclosed mass at $200M$, but they build this mass at different rates. This profile dependence also appears in the seed function, the time conversion, and the dominant tidal eigenvalue.

\section{The Intrinsic Intermediate-Axis Clock}
\label{sec:intrinsic}

\subsection{Free triaxial rotation}

For principal moments $I_1<I_2<I_3$, free rotation obeys the Euler equations
\begin{align}
 I_1\dot\omega_1&=(I_2-I_3)\omega_2\omega_3,\nonumber\\
 I_2\dot\omega_2&=(I_3-I_1)\omega_3\omega_1,\label{eq:eulerfree}\\
 I_3\dot\omega_3&=(I_1-I_2)\omega_1\omega_2.\nonumber
\end{align}
Rotation about the first and third principal axes is stable. Rotation about the second axis is unstable. A small transverse component grows and leads to repeated sign changes and orientation reversals \cite{Goldstein:2002,Landau:1976,VanDamme:2017,Mardesic:2020}.

The free system conserves
\begin{equation}
 E_{\rm rot}=\frac12\sum_i I_i\omega_i^2,
 \qquad L^2=\sum_i I_i^2\omega_i^2.
\label{eq:invariants}
\end{equation}
These two quantities are used as direct code checks. The motion also has a geometric phase in the reconstruction of the spatial orientation \cite{Montgomery:1991}. The quantity of interest is therefore the full orientation history, rather than the body-frame angular velocity alone.

The absolute flip period changes strongly with the moments and with the distance from the unstable separatrix. The absolute period cannot be predicted from the environment alone. I therefore compare each DM run with a Kerr run that has exactly the same body parameters.

\subsection{Reduced quadrupolar element}

The element has a trace-free body-frame quadrupole
\begin{equation}
 Q_{ij}^{\rm body}=\mathrm{diag}(Q_1,Q_2,Q_3),
 \qquad Q_1+Q_2+Q_3=0.
\label{eq:quadrupole}
\end{equation}
Its orientation is represented by a unit quaternion $q$. If $R(q)$ maps body coordinates to the local tetrad, then
\begin{equation}
 E^{\rm body}=R(q)^{\mathsf T}E^{\rm local}R(q).
\label{eq:bodytidal}
\end{equation}
The reduced tidal torque is
\begin{equation}
 N_i^{\rm tidal}=K_{\rm tidal}\epsilon_{ijk}Q_{jl}E^{\rm body}_{kl}.
\label{eq:torque}
\end{equation}
The dimensionless coefficient $K_{\rm tidal}$ is an effective sensitivity. It is not a measured material constant.

The proper-time equations are
\begin{align}
 I_1\frac{\dd\omega_1}{\dd\tau}&=(I_2-I_3)\omega_2\omega_3+N_1^{\rm tidal},\nonumber\\
 I_2\frac{\dd\omega_2}{\dd\tau}&=(I_3-I_1)\omega_3\omega_1+N_2^{\rm tidal},\label{eq:eulerrel}\\
 I_3\frac{\dd\omega_3}{\dd\tau}&=(I_1-I_2)\omega_1\omega_2+N_3^{\rm tidal},\nonumber
\end{align}
and the quaternion convention is
\begin{equation}
 \frac{\dd q}{\dd\tau}=\frac12 q\otimes(0,\vect{\omega}),
 \qquad q_Aq_A=1.
\label{eq:quat}
\end{equation}
This model retains the local quadrupole--curvature structure of extended-body dynamics \cite{Dixon:1974c,EhlersRudolph:1977,SteinhoffPuetzfeld:2012}, but it is not the complete Mathisson--Papapetrou--Dixon system. The orbit, spin supplementary condition, and material quadrupole are not evolved covariantly.

\section{Local Tidal Coupling and the Orientation Clock}
\label{sec:clock}

\subsection{Circular orbit and tidal tensor}

For a stationary axisymmetric metric, the prograde angular velocity of a circular equatorial geodesic is
\begin{equation}
 \Omega_\phi=\frac{-\partial_r g_{t\phi}+
 \sqrt{(\partial_r g_{t\phi})^2-(\partial_r g_{tt})(\partial_r g_{\phi\phi})}}
 {\partial_r g_{\phi\phi}}.
\label{eq:omega}
\end{equation}
The time component of the four-velocity is
\begin{equation}
 u^t=\left[-g_{tt}-2\Omega_\phi g_{t\phi}
 -\Omega_\phi^2g_{\phi\phi}\right]^{-1/2}.
\label{eq:ut}
\end{equation}
A Gram--Schmidt construction gives a comoving tetrad $e_{\hat\alpha}^{\ \mu}$ with
\begin{equation}
 g_{\mu\nu}e_{\hat\alpha}^{\ \mu}e_{\hat\beta}^{\ \nu}
 =\eta_{\hat\alpha\hat\beta}.
\label{eq:ortho}
\end{equation}
I use the curvature convention
\begin{equation}
 R^{\rho}{}_{\sigma\mu\nu}
 =\partial_\mu\Gamma^{\rho}_{\nu\sigma}
 -\partial_\nu\Gamma^{\rho}_{\mu\sigma}
 +\Gamma^{\rho}_{\mu\lambda}\Gamma^{\lambda}_{\nu\sigma}
 -\Gamma^{\rho}_{\nu\lambda}\Gamma^{\lambda}_{\mu\sigma}.
\label{eq:riemannconv}
\end{equation}
The electric tidal tensor measured by the orbiting observer is
\begin{equation}
 E_{\hat i\hat j}=R_{\mu\alpha\nu\beta}
 e_{\hat i}^{\ \mu}u^\alpha e_{\hat j}^{\ \nu}u^\beta.
\label{eq:tidal}
\end{equation}
It is symmetric. In vacuum Kerr its trace vanishes, up to numerical differentiation error. The construction is closely related to geodesic deviation and to the tidal tensors obtained in parallel-propagated Kerr frames \cite{ManasseMisner:1963,Marck:1983}.

\subsection{Operational clock definition}

The event detector records every zero crossing of $\omega_2$, in both directions. Adjacent crossings are half-period markers. I discard the first interval to reduce sensitivity to the prepared initial state, and I define the full proper-time clock period by
\begin{equation}
 T_{\flip}^{(\tau)}=2\left\langle\tau_{n+1}-\tau_n\right\rangle,
 \qquad \nu_{\flip}^{(\tau)}=\frac{1}{T_{\flip}^{(\tau)}}.
\label{eq:propertimeclock}
\end{equation}
Adjacent crossings are half a cycle apart, so their mean separation is doubled. The reported period is one full orientation cycle.

For a circular orbit, $u^t$ is constant during each reduced integration. Therefore
\begin{equation}
 T_{\flip}^{(t)}=u^tT_{\flip}^{(\tau)},\qquad
 \nu_{\flip}^{(t)}=\frac{\nu_{\flip}^{(\tau)}}{u^t}.
\label{eq:clock}
\end{equation}
I define
\begin{align}
 \delta_\tau&=\frac{\nu_{\flip,\DM}^{(\tau)}}
 {\nu_{\flip,\Kerr}^{(\tau)}}-1,\nonumber\\
 \delta_{u^t}&=\frac{u^t_{\Kerr}}{u^t_{\DM}}-1,\label{eq:deltas}\\
 \delta_t&=\frac{\nu_{\flip,\DM}^{(t)}}
 {\nu_{\flip,\Kerr}^{(t)}}-1.\nonumber
\end{align}
Equation~\eqref{eq:clock} then gives the exact identity
\begin{equation}
 1+\delta_t=(1+\delta_\tau)(1+\delta_{u^t}).
\label{eq:factorization}
\end{equation}
To first order, $\delta_t\simeq\delta_\tau+\delta_{u^t}$.

Equation~\eqref{eq:factorization} organizes the analysis. It follows directly from Eq.~\eqref{eq:clock} and separates the local dynamical response from the time-conversion response. The first factor changes the local orientation dynamics. The second changes the relation between the orbiting clock and the stationary coordinate time. When $K_{\rm tidal}=0$, the first term vanishes but the second remains. The numerical calculation then determines which channel dominates for each body model and environment.

\begin{figure}[!t]
\centering
\includegraphics[width=\textwidth]{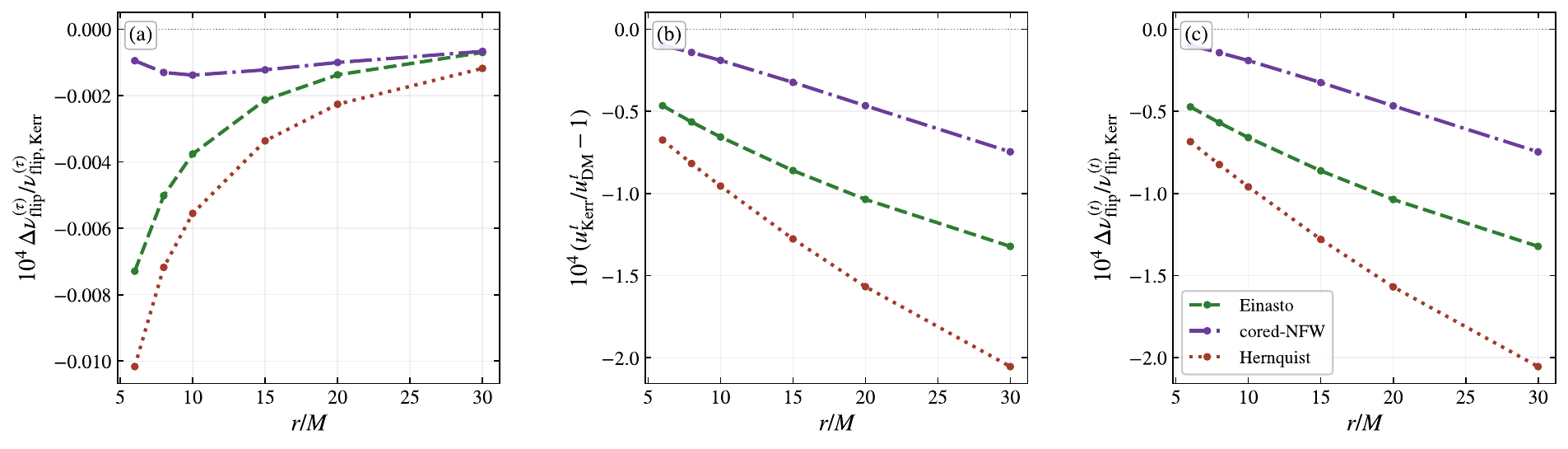}
\caption{Direct clock decomposition for $\chi=0.8$, $\mu_{\DM}=0.05$, and $K_{\rm tidal}=1$. The panels show the proper-time response, the time-conversion response, and the exact total coordinate-time shift. The vertical scale is $10^4$ times the fractional frequency change.}
\label{fig:decomposition}
\end{figure}

Figure~\ref{fig:decomposition} shows the direct calculation. For the fiducial body, the local proper-time response is small and the time-conversion term controls the total ordering of the three profiles.

\FloatBarrier

\section{Numerical Strategy and Validation}
\label{sec:numerics}

\subsection{Overview of the second half of the paper}

{The first four sections define the geometry, the local tidal field, and the orientation clock. From this point, the paper has three tasks. This section validates the numerical chain. Sections~\ref{sec:profiles}--\ref{sec:time} then map the response to the halo and body models and follow the same solutions in time. Sections~\ref{sec:forecast} and \ref{sec:context} translate that response into a synthetic timing test and physical units. No new geometry is introduced after Sec.~\ref{sec:clock}; every later result is an application of the same clock decomposition in Eq.~\eqref{eq:factorization}.}

\subsection{Direct calculation and validation}

The direct calculation follows one fixed chain,
\begin{equation}
 \rho_{\DM}\longrightarrow M_{\DM}\longrightarrow g_{\mu\nu}
 \longrightarrow (u^t,E_{\hat i\hat j})
 \longrightarrow (q,\vect\omega)
 \longrightarrow \nu_{\flip}^{(t)}.
\label{eq:pipeline}
\end{equation}
{Profile integrals use monotone piecewise-cubic interpolation \cite{FritschCarlson:1980}. Metric derivatives and curvature use centered finite differences with radius-dependent steps. I integrate the Euler--quaternion system with the adaptive DOP853 Runge--Kutta method \cite{Hairer:1993} and locate flip times with event detection. The first half-period is removed as a transient, and the mean of the remaining half-periods is doubled as in Eq.~\eqref{eq:propertimeclock}. Each DM run is paired with a Kerr run that uses the same body and numerical settings. The implementation uses NumPy, SciPy, and Matplotlib \cite{Harris:2020xlr,Virtanen:2019joe,Hunter:2007}.}

{The production database contains 600 DM runs and 40 paired Kerr references. The grid uses $\chi\in\{0,0.5,0.8,0.95\}$, $r/M\in\{6,7,8,10,12,15,20,30,40,60\}$, and $\mu_{\DM}\in\{0.002,0.005,0.01,0.02,0.05\}$. The baseline body has $(I_1,I_2,I_3)=(1,2,3)$, $\vect\omega(0)=(0.04,1,0.02)$, and $K_{\rm tidal}=1$.}

{Table~\ref{tab:validation} summarizes the main numerical checks. Kerr orbital motion is recovered to better than $6\times10^{-9}$. Tetrad orthonormality and the exact clock identity are satisfied near machine precision. The Kerr tidal trace checks the curvature derivatives, while the free-top energy, angular momentum, and quaternion norm check the internal solver. Higher resolution changes the total environmental shift by at most $4.04\times10^{-4}$ relative to the baseline value. The separate horizon scan finds no additional outer zero of $\Delta$ in any of the 60 background geometries.}

\begin{table}[!t]
\centering
\caption{Validation of the direct production calculation.}
\label{tab:validation}
\begin{tabular}{@{}lr@{}}
\toprule
Diagnostic & Result or bound \\
\midrule
Kerr orbital-frequency relative error & $5.01\times10^{-9}$ \\
Tetrad orthonormality error & $4.45\times10^{-16}$ \\
Relative Kerr tidal trace & $4.27\times10^{-8}$ \\
Additional outer zeros of $\Delta$ (60 backgrounds) & 0 \\
Minimum $\partial_r\Delta$ at $r_+^{\rm K}$ & $6.24\times10^{-1}$ \\
Clock-factorization residual & $2.23\times10^{-16}$ \\
Crossing-interval scatter fraction & $4.00\times10^{-4}$ \\
High-resolution total-shift difference & $4.04\times10^{-4}$ \\
Quaternion norm error & $1.85\times10^{-10}$ \\
Free-top energy drift & $8.38\times10^{-10}$ \\
Free-top $L^2$ drift & $8.38\times10^{-10}$ \\
\bottomrule
\end{tabular}
\end{table}

{In the baseline model, the proper-time channel is much smaller than the total shift. Its relative convergence error can reach $4.3$ per cent even though its absolute error is very small. I therefore take all central values from the direct grid.}

\begin{figure}[!t]
\centering
\includegraphics[width=0.94\textwidth]{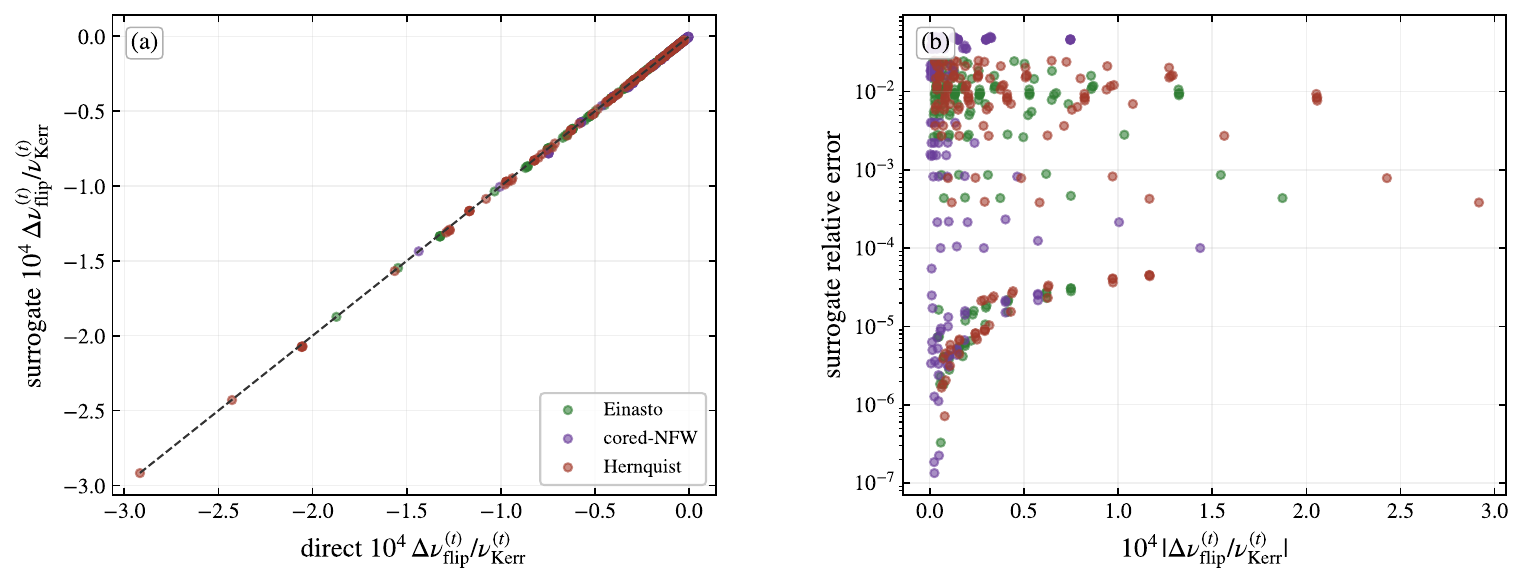}
\caption{Cross-validation of the bounded surrogate. The two clock channels are interpolated separately and joined with the exact identity in Eq.~\eqref{eq:factorization}. The right panel shows the relative error in the total shift. Direct calculations remain the source of the reported central values.}
\label{fig:surrogate}
\end{figure}

{For dense forecast maps only, I use a fast surrogate. It interpolates $\delta_\tau$ and $\delta_{u^t}$ separately and then reconstructs $\delta_t$ with Eq.~\eqref{eq:factorization}. Test points are excluded from the coarse training grid. The median relative errors in the total shift are $0.8$, $1.8$, and $0.8$ per cent for Einasto, cored-NFW, and Hernquist. The 95th percentiles are $2.3$, $4.9$, and $2.4$ per cent. Figure~\ref{fig:surrogate} shows the cross-validation.}

\section{Profile Dependence of the Clock}
\label{sec:profiles}

\begin{figure}[!t]
\centering
\includegraphics[width=\textwidth]{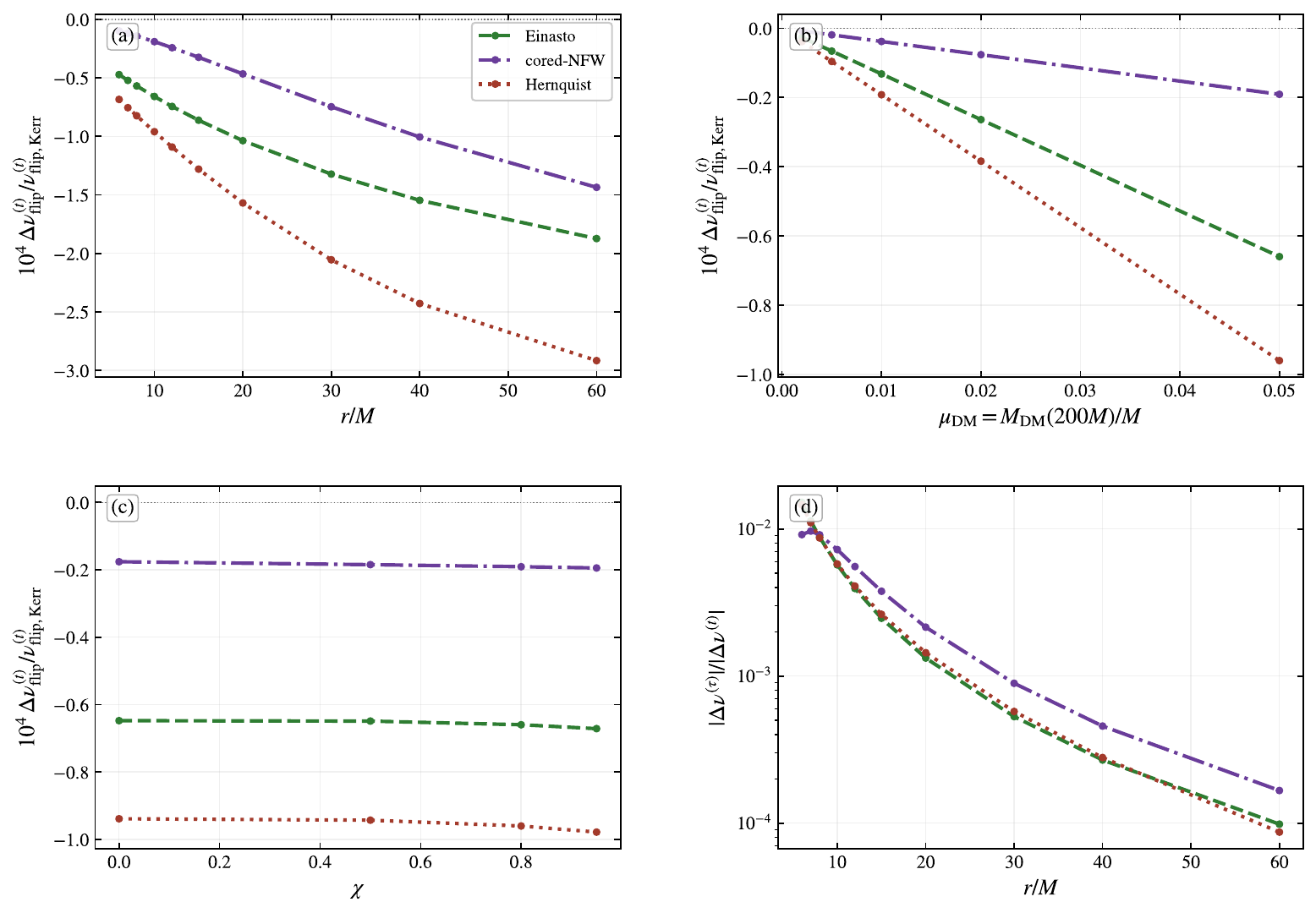}
\caption{Direct production grid. (a) Radius dependence for $\chi=0.8$ and $\mu_{\DM}=0.05$. (b) Halo-normalization dependence at $r=10M$ and $\chi=0.8$. (c) Spin dependence at $r=10M$ and $\mu_{\DM}=0.05$. (d) Fraction of the total shift carried by the proper-time response.}
\label{fig:grid}
\end{figure}

{With the numerical chain validated, I first vary the environment while keeping the reduced body fixed. This isolates how the radial mass distribution changes the two clock channels. Figure~\ref{fig:grid} summarizes the direct grid. Similar environmental calculations show that strong-field observables depend on the radial distribution of the surrounding matter, not only on one total mass value \cite{Barausse:2014tra,Kavanagh:2020cfn,Cardoso:2021wlq,Figueiredo:2023gas}. At fixed $\chi=0.8$ and $\mu_{\DM}=0.05$, the magnitude of the shift grows with radius because the enclosed halo mass continues to grow. I report this trend only within the model domain.}

The shift is almost linear in $\mu_{\DM}$ over the scanned range. At fixed profile, spin, and radius, the median fractional residual of a linear fit is $1.7\times10^{-5}$, and the largest residual is $3.5\times10^{-4}$. Spin changes the shift less than radius and halo normalization in this grid.

Representative values are listed in Table~\ref{tab:shifts}. All listed shifts are negative: in this model the DM-dressed coordinate-time frequency is slightly smaller than the paired Kerr frequency. For the adopted shapes and common normalization, the usual ordering is
\begin{equation}
 |\delta_t|_{\rm H}>|\delta_t|_{\rm E}>|\delta_t|_{\cNFW}.
\label{eq:ordering}
\end{equation}
This ordering is specific to the three profile shapes and the common normalization used here.

\begin{table}[!t]
\centering
\caption{Direct clock shifts for $\chi=0.8$, $\mu_{\DM}=0.05$, and $K_{\rm tidal}=1$.}
\label{tab:shifts}
\setlength{\tabcolsep}{3.5pt}
\begin{tabular}{@{}lrrrr@{}}
\toprule
Profile & $r/M$ & $\delta_\tau$ & $\delta_{u^t}$ & $\delta_t$ \\
\midrule
Einasto & 6  & $-7.16\!\times\!10^{-7}$ & $-4.66\!\times\!10^{-5}$ & $-4.73\!\times\!10^{-5}$ \\
        & 10 & $-3.76\!\times\!10^{-7}$ & $-6.56\!\times\!10^{-5}$ & $-6.60\!\times\!10^{-5}$ \\
        & 20 & $-1.37\!\times\!10^{-7}$ & $-1.04\!\times\!10^{-4}$ & $-1.04\!\times\!10^{-4}$ \\
        & 60 & $-1.84\!\times\!10^{-8}$ & $-1.87\!\times\!10^{-4}$ & $-1.87\!\times\!10^{-4}$ \\
\addlinespace
cored-NFW & 6  & $-9.08\!\times\!10^{-8}$ & $-9.83\!\times\!10^{-6}$ & $-9.92\!\times\!10^{-6}$ \\
          & 10 & $-1.38\!\times\!10^{-7}$ & $-1.90\!\times\!10^{-5}$ & $-1.91\!\times\!10^{-5}$ \\
          & 20 & $-1.00\!\times\!10^{-7}$ & $-4.65\!\times\!10^{-5}$ & $-4.66\!\times\!10^{-5}$ \\
          & 60 & $-2.39\!\times\!10^{-8}$ & $-1.44\!\times\!10^{-4}$ & $-1.44\!\times\!10^{-4}$ \\
\addlinespace
Hernquist & 6  & $-9.98\!\times\!10^{-7}$ & $-6.74\!\times\!10^{-5}$ & $-6.84\!\times\!10^{-5}$ \\
          & 10 & $-5.55\!\times\!10^{-7}$ & $-9.55\!\times\!10^{-5}$ & $-9.60\!\times\!10^{-5}$ \\
          & 20 & $-2.26\!\times\!10^{-7}$ & $-1.57\!\times\!10^{-4}$ & $-1.57\!\times\!10^{-4}$ \\
          & 60 & $-2.53\!\times\!10^{-8}$ & $-2.92\!\times\!10^{-4}$ & $-2.92\!\times\!10^{-4}$ \\
\bottomrule
\end{tabular}
\end{table}

Across the full grid, $|\delta_\tau|/|\delta_t|$ has a median of $3.8\times10^{-3}$ and a maximum of $2.13\times10^{-2}$. The time-conversion dominance therefore emerges from the fiducial calculation rather than being imposed in advance.

\section{Matter-Model Robustness}
\label{sec:robustness}

{The profile scan fixes the body model. I now do the opposite: I vary the reduced body while keeping each DM/Kerr comparison paired. This tests whether the environmental shift depends strongly on one prepared triaxial element. The intrinsic Kerr clock is expected to depend on the body because of the classical separatrix structure and curvature--multipole coupling \cite{Montgomery:1991,Dixon:1974c,SteinhoffPuetzfeld:2012}. I vary the tidal sensitivity, inertia ratios, transverse seed, initial orientation, and profile scale.}

\begin{figure}[!t]
\centering
\includegraphics[width=\textwidth]{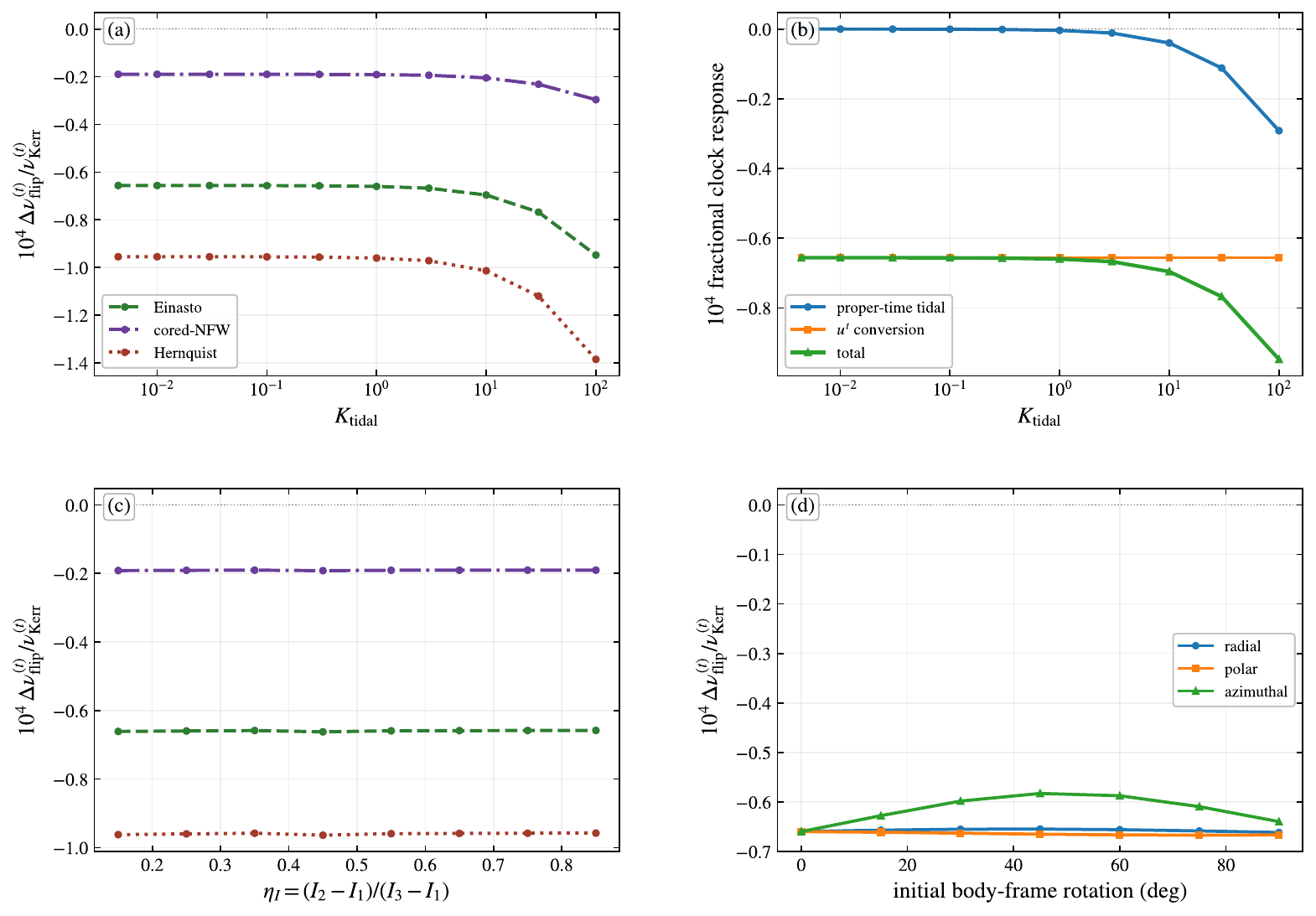}
\caption{Robustness to the reduced-body model. The panels vary the tidal sensitivity, the inertia ratio, the transverse seed, and the initial body orientation. The absolute intrinsic frequency may change strongly, but the paired fractional environmental response remains much more stable near $K_{\rm tidal}=1$.}
\label{fig:robustness}
\end{figure}

Figure~\ref{fig:robustness} gives the main scans for $\chi=0.8$, $r=10M$, and $\mu_{\DM}=0.05$. At $K_{\rm tidal}=0$, $\delta_\tau=0$ exactly, while $\delta_t=\delta_{u^t}$ remains. Between $K_{\rm tidal}=0$ and $1$, the total shift changes only slightly. At the extreme value $K_{\rm tidal}=100$, the proper-time term becomes visible. The total shifts become $-9.48\times10^{-5}$, $-2.96\times10^{-5}$, and $-1.39\times10^{-4}$ for Einasto, cored-NFW, and Hernquist. The $K_{\rm tidal}=100$ case is a stress test of a very strong effective coupling.

The inertia scan changes the intrinsic Kerr frequency strongly, but the total DM shift stays within about one per cent of its profile-specific mean. In the Einasto seed map, the Kerr frequency ranges from $2.16\times10^{-2}$ to $5.74\times10^{-2}$ in geometrized units. Over the same map, the total shift remains between $-6.60\times10^{-5}$ and $-6.58\times10^{-5}$. Much of the preparation dependence cancels in the paired comparison because $\delta_{u^t}$ does not depend on the body.

\begin{figure}[!t]
\centering
\includegraphics[width=0.94\textwidth]{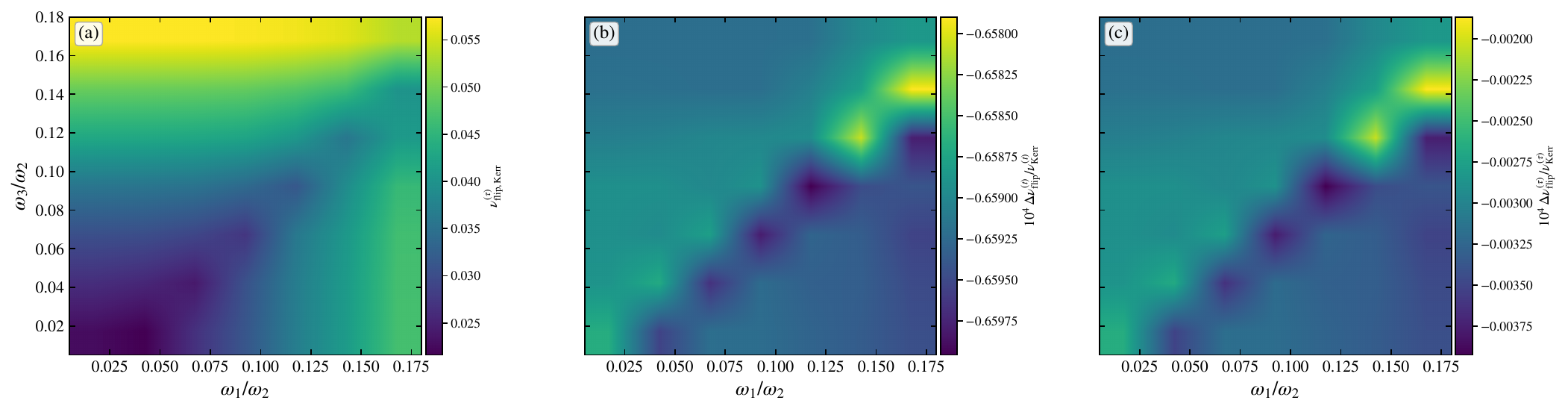}
\caption{Direct seed-component scan. The intrinsic Kerr clock changes strongly close to the unstable separatrix. The paired Kerr/DM fractional response remains smooth over the retained region.}
\label{fig:seedmap}
\end{figure}

\begin{figure}[!t]
\centering
\includegraphics[width=0.94\textwidth]{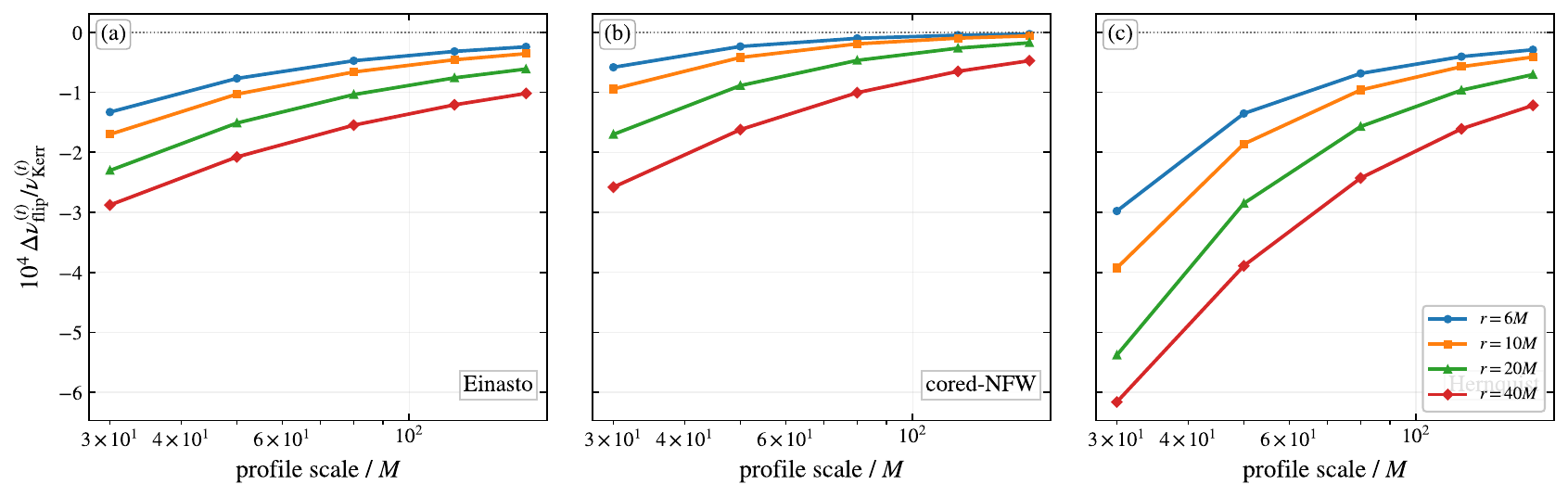}
\caption{Profile-scale scan at fixed $M_{\DM}(200M)$. The response depends on the radial buildup of the enclosed mass. Fixing the mass at one outer radius does not remove this profile dependence.}
\label{fig:profilescale}
\end{figure}

Figure~\ref{fig:seedmap} extends the seed scan over both transverse components. The intrinsic frequency changes rapidly near the separatrix, while the fractional DM response is smooth. Figure~\ref{fig:profilescale} changes the profile scale at fixed $M_{\DM}(200M)$. The clock responds to where the mass is placed, not only to the mass at the normalization radius.

\section{Time-Domain Dynamics and Projected Morphology}
\label{sec:time}

\begin{figure}[!t]
\centering
\includegraphics[width=\textwidth]{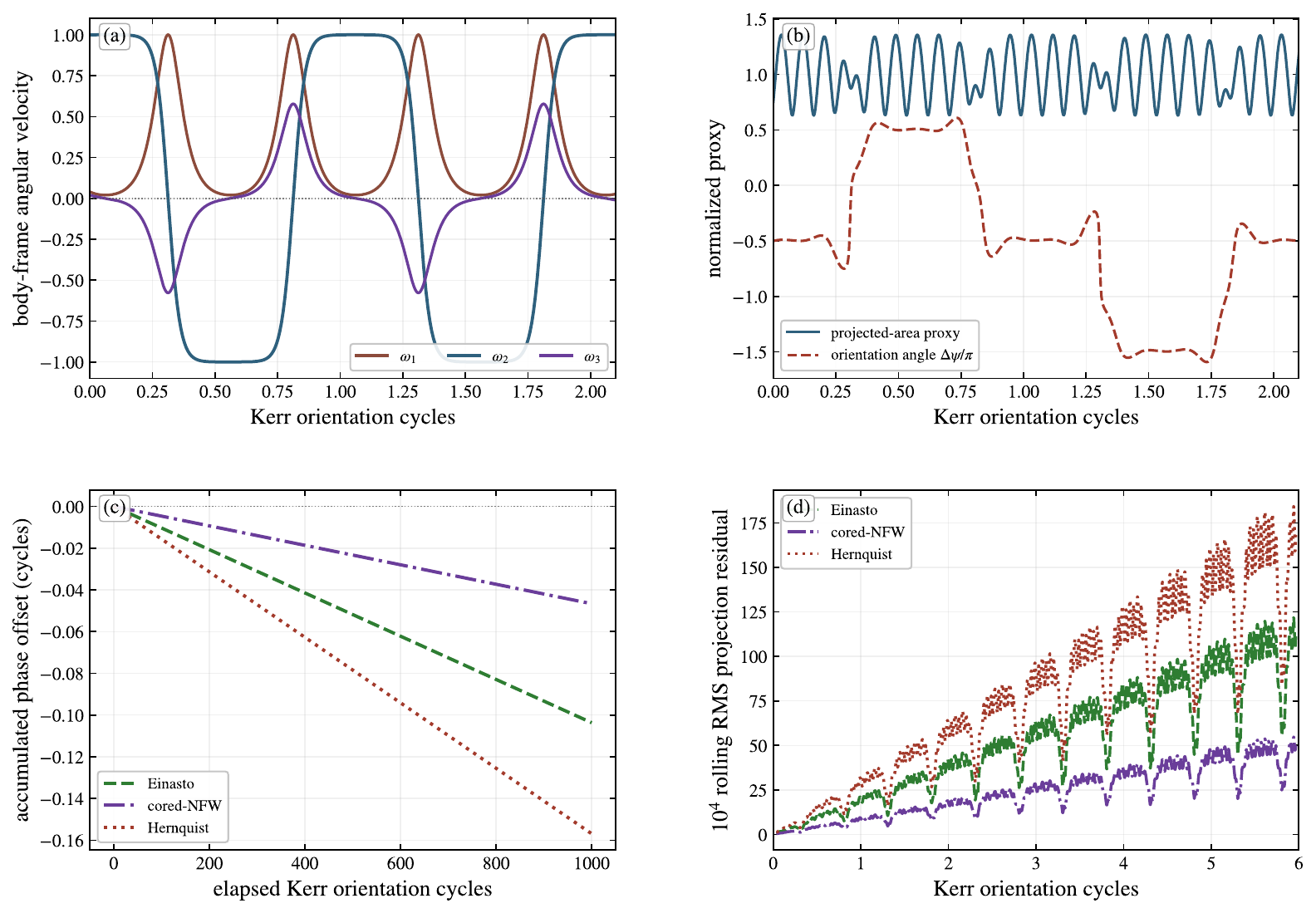}
\caption{Direct time-domain evolution. (a) Kerr body-frame angular velocities. (b) Kerr projected-area proxy and an orientation-angle diagnostic. (c) Ideal coherent phase drift obtained from the directly measured profile frequencies. The extension to 1000 cycles is an ideal phase-accumulation test. (d) Rolling root-mean-square difference between each profile proxy and Kerr over the directly integrated interval.}
\label{fig:time}
\end{figure}

{Sections~\ref{sec:profiles} and \ref{sec:robustness} establish the size and stability of the frequency shift. I now follow the same solutions in time to show what this small shift means for the orientation itself. The full quaternion system is integrated at common coordinate times for Kerr and for the three profiles. Quaternion reconstruction keeps the complete spatial orientation, including the geometric phase of triaxial rotation \cite{Montgomery:1991}. Figure~\ref{fig:time} shows the case $\chi=0.8$, $r=20M$, $\mu_{\DM}=0.05$, and $K_{\rm tidal}=1$. The body-frame angular velocities show the expected intermediate-axis reversals. A simple projected-area diagnostic changes with the orientation. Over one or two cycles, the DM curves are almost indistinguishable from Kerr. Their phase difference grows slowly and systematically.}

After 1000 ideal Kerr orientation-clock cycles, the phase offsets are $-0.104$, $-0.0466$, and $-0.157$ cycles for Einasto, cored-NFW, and Hernquist. This shows how a small frequency shift can build a measurable phase only if a stable reference survives for many cycles. The reduced model does not prove such long coherence.

\begin{figure}[!t]
\centering
\includegraphics[width=0.94\textwidth]{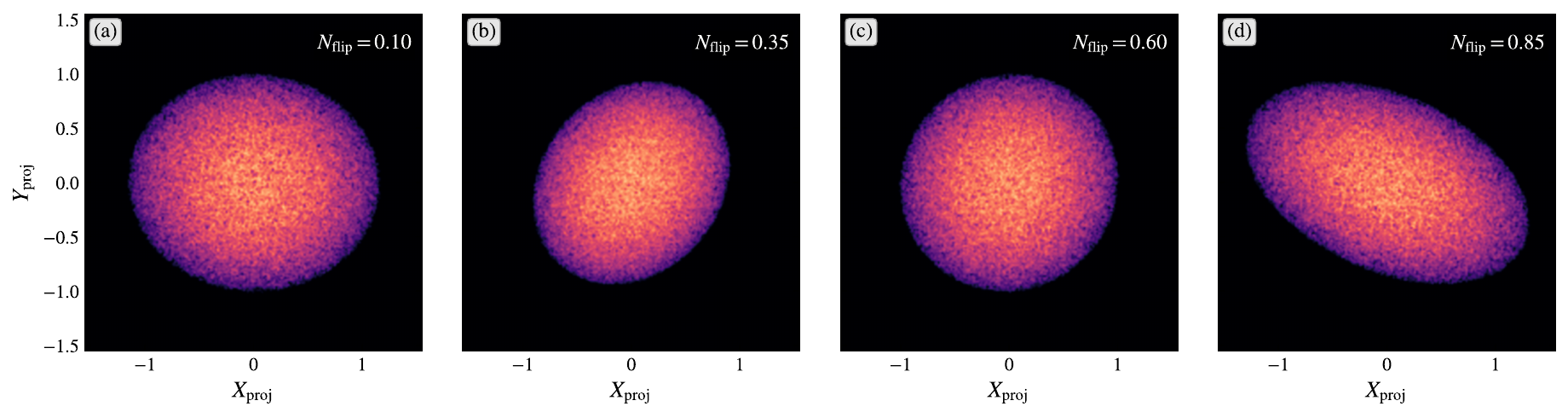}
\caption{Orthographic projection-plane tracer-emissivity proxy at four phases of the direct orientation history. The same three-dimensional cloud is used in every panel. Changes in apparent size and shape are caused only by projection.}
\label{fig:projection}
\end{figure}

To illustrate the orientation geometry, I rotate one fixed three-dimensional triaxial tracer cloud with the direct quaternion history and project it orthographically. Figure~\ref{fig:projection} uses the coordinates $X_{\rm proj}$ and $Y_{\rm proj}$. The apparent stretching and shrinking come from the changing projection of the same cloud. The brightness is a weighted tracer-emissivity proxy. This construction includes no lensing, GRMHD evolution, or radiative transfer.

\section{Synthetic Timing Forecast}
\label{sec:forecast}

\subsection{Scope of the synthetic forecast}

{The previous sections give a geometric frequency shift. I now ask how difficult it would be to recover that shift from a simple timing signal. The relativistic calculation supplies the frequency change; the timing model adds modulation, noise, cadence, and finite coherence. Published measurements are used only to set the scale. No real source is fitted.}

I use the simple modulation
\begin{equation}
 F(t)=F_0\left[1+A_1\cos(2\pi f t+\phi_1)
 +A_2\cos(4\pi f t+\phi_2)\right]+n(t),
\label{eq:flux}
\end{equation}
where $n(t)$ contains white and autoregressive red noise. The model contains a fundamental and a first harmonic of an orientation signal. It is deliberately simpler than radiative transfer. The frequency is estimated by prewhitened harmonic regression. Lomb--Scargle methods provide the natural extension to irregular sampling \cite{Lomb:1976,Scargle:1982,VanderPlas:2018}, while red-noise significance requires separate care \cite{Vaughan:2005,BachettiHuppenkothen:2024}.

\begin{figure}[!t]
\centering
\includegraphics[width=\textwidth]{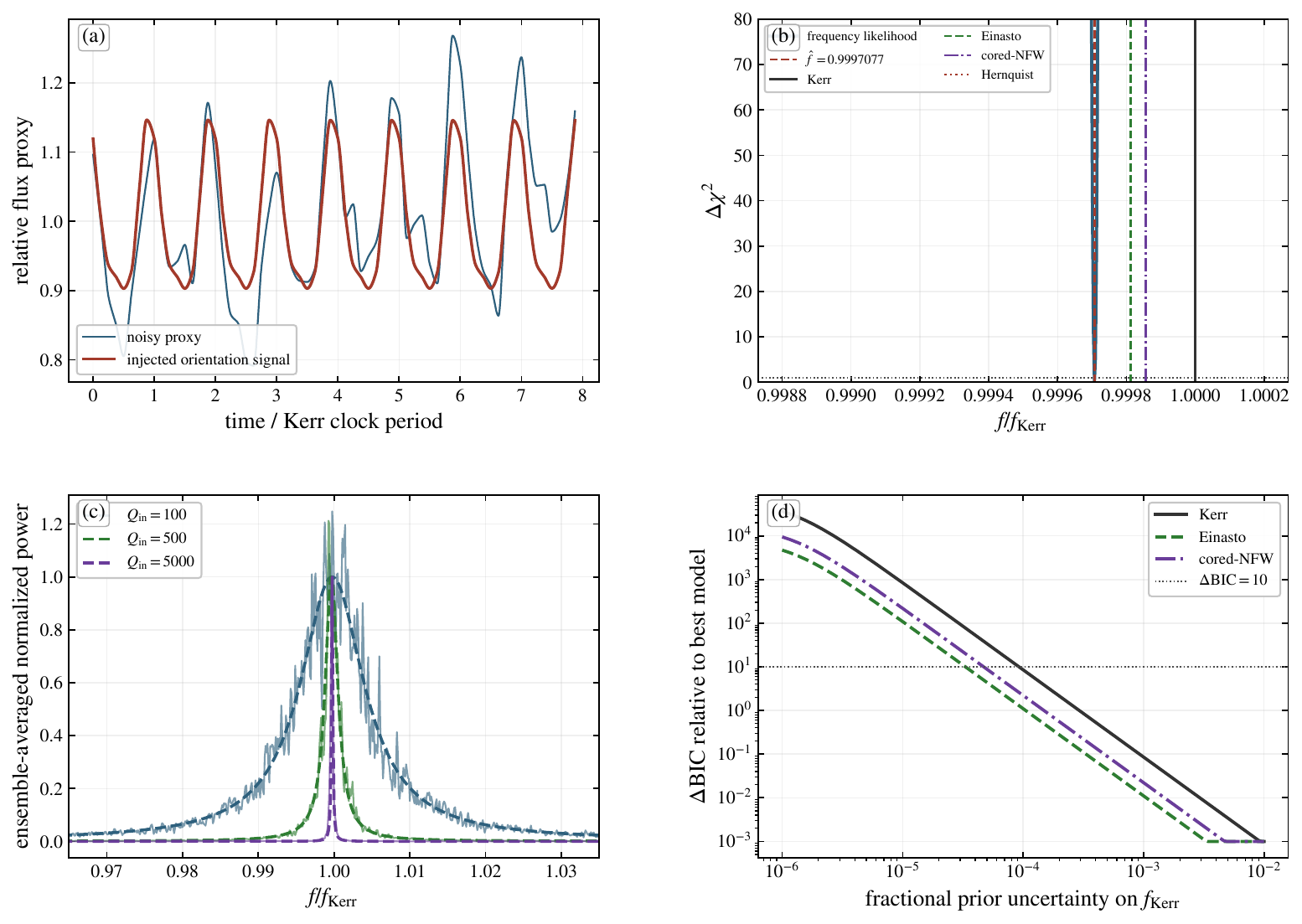}
\caption{Synthetic timing recovery. (a) A short part of the noisy orientation proxy. (b) Frequency likelihood and fixed profile predictions when the Kerr baseline is known. (c) Ensemble-averaged phase-diffusion spectra; dashed lines are Lorentzian fits. (d) Loss of model separation as the prior uncertainty on the Kerr clock increases. The injected example is deliberately coherent and has high signal-to-noise ratio.}
\label{fig:synthetic}
\end{figure}

The recovery test injects the largest shift in the $\chi=0.8$, $\mu_{\DM}=0.05$ slice: the Hernquist case at $60M$,
\begin{equation}
 \frac{f_{\rm inj}}{f_{\Kerr}}=1-2.916\times10^{-4}.
\label{eq:injection}
\end{equation}
The clock lasts for 2000 cycles with eight samples per cycle. The recovered frequency is $0.99970768f_{\Kerr}$, compared with the injected value $0.99970841f_{\Kerr}$. The local likelihood curvature gives $\sigma_f/f\simeq1.15\times10^{-6}$. This deliberately favorable setup tests the inference pipeline under controlled conditions. {Figure~\ref{fig:synthetic} collects the recovery, coherence, and baseline-degeneracy diagnostics described here and in the next two subsections.}

\subsection{Finite coherence}

I model finite coherence by phase diffusion. If $D$ is the phase-diffusion coefficient, I set $D=\pi f/Q$, where $Q$ is the quality factor. Ensemble-averaged Welch spectra \cite{Welch:1967} recover $Q=97.3$, $494.5$, and $5000.2$ for injected values $100$, $500$, and $5000$. The timing pipeline therefore reproduces the intended broadening.

For a simple sinusoidal clock, an optimistic lower bound is
\begin{equation}
 \frac{\sigma_f}{f}\simeq\frac{\sqrt{3}}{\pi\rho N_{\rm eff}},
 \qquad N_{\rm eff}=\min\left(N_{\rm cyc},\frac{Q}{\pi}\right),
\label{eq:sigmaf}
\end{equation}
where $\rho$ is the total signal-to-noise ratio and $N_{\rm cyc}$ is the observed number of cycles.

\subsection{The baseline degeneracy}

If all profile frequencies are fixed and the Kerr baseline is known, the models can be compared with $\chi^2$ or the Bayesian information criterion (BIC) \cite{Schwarz:1978}. In the favorable injection, the fixed-frequency $\Delta{\rm BIC}$ values relative to Hernquist are $1.84\times10^4$, $7.27\times10^3$, and $1.12\times10^4$ for Kerr, Einasto, and cored-NFW.\@ These large numbers reflect the deliberately favorable calibration of the synthetic example.

{The important limitation is the degeneracy that appears when the intrinsic Kerr frequency $f_0$ is free.} Each profile predicts $f=(1+\delta_p)f_0$, so one measured frequency can always be absorbed into a slightly different $f_0$. I therefore add a Gaussian prior and minimize
\begin{equation}
 \chi_p^2=\frac{[\hat f-(1+\delta_p)f_0]^2}{\sigma_f^2}
 +\frac{(f_0-f_{0,{\rm prior}})^2}{\sigma_{f_0}^2}.
\label{eq:prior}
\end{equation}
Figure~\ref{fig:synthetic}(d) shows that profile separation disappears when the baseline uncertainty becomes comparable with the environmental shift.

\begin{figure}[!t]
\centering
\includegraphics[width=\textwidth]{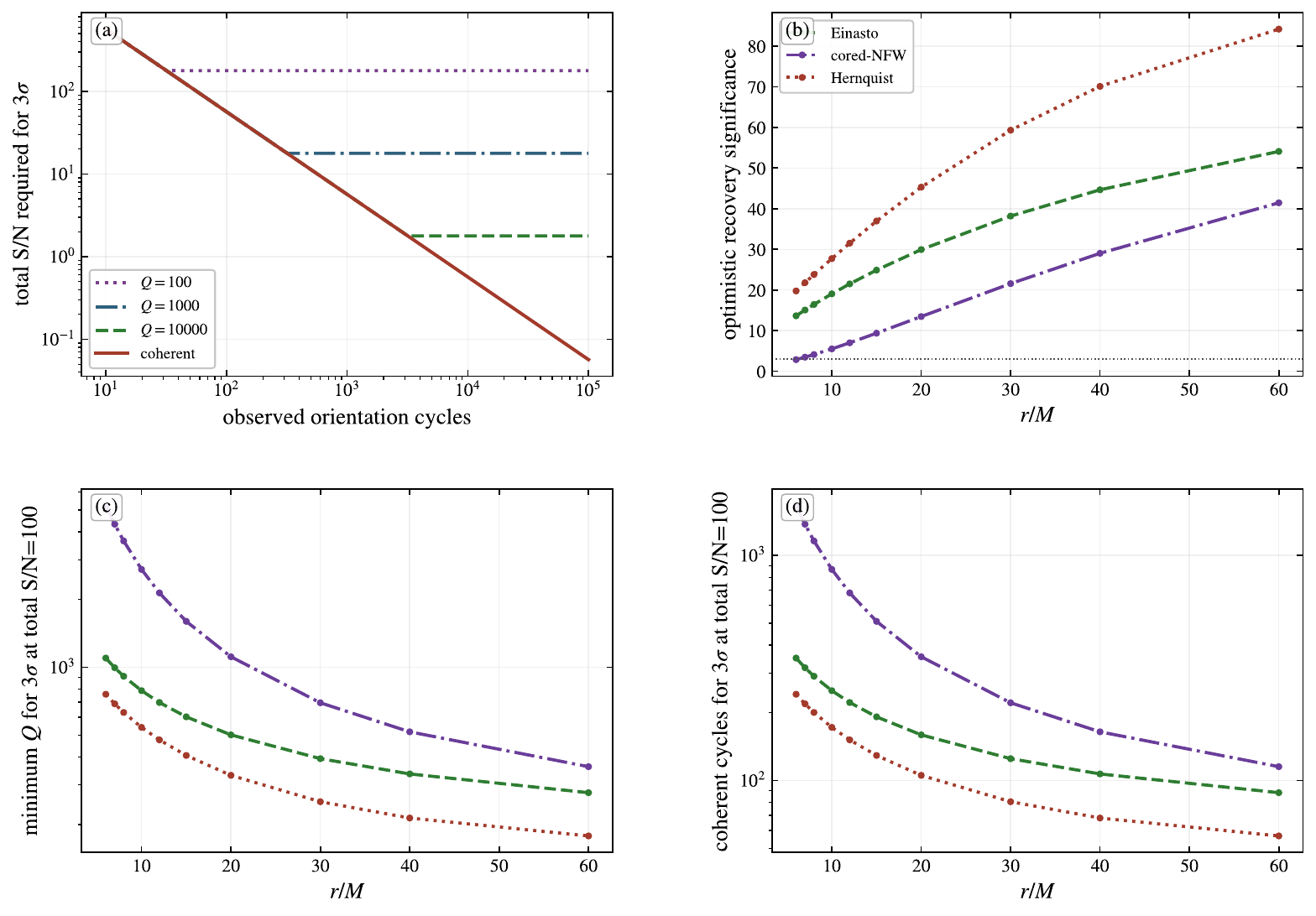}
\caption{Optimistic detectability forecast. (a) Total signal-to-noise ratio needed for a $3\sigma$ recovery of the fiducial Hernquist shift. (b) Profile significance for one representative setup. (c) Minimum quality factor and (d) minimum coherent cycle count at total signal-to-noise ratio 100. These are lower bounds from Eq.~\eqref{eq:sigmaf}.}
\label{fig:detect}
\end{figure}

Figure~\ref{fig:detect} gives an optimistic detectability map based on Eq.~\eqref{eq:sigmaf}. Broad quasi-periodic oscillations (QPOs) with moderate $Q$ are not expected to resolve a $10^{-4}$ shift. A better target would be a repeated clock whose Kerr baseline is constrained by other measurements, or a differential change followed in the same source.

\section{Physical Scaling and Established Timing Channels}
\label{sec:context}

\begin{figure}[!t]
\centering
\includegraphics[width=\textwidth]{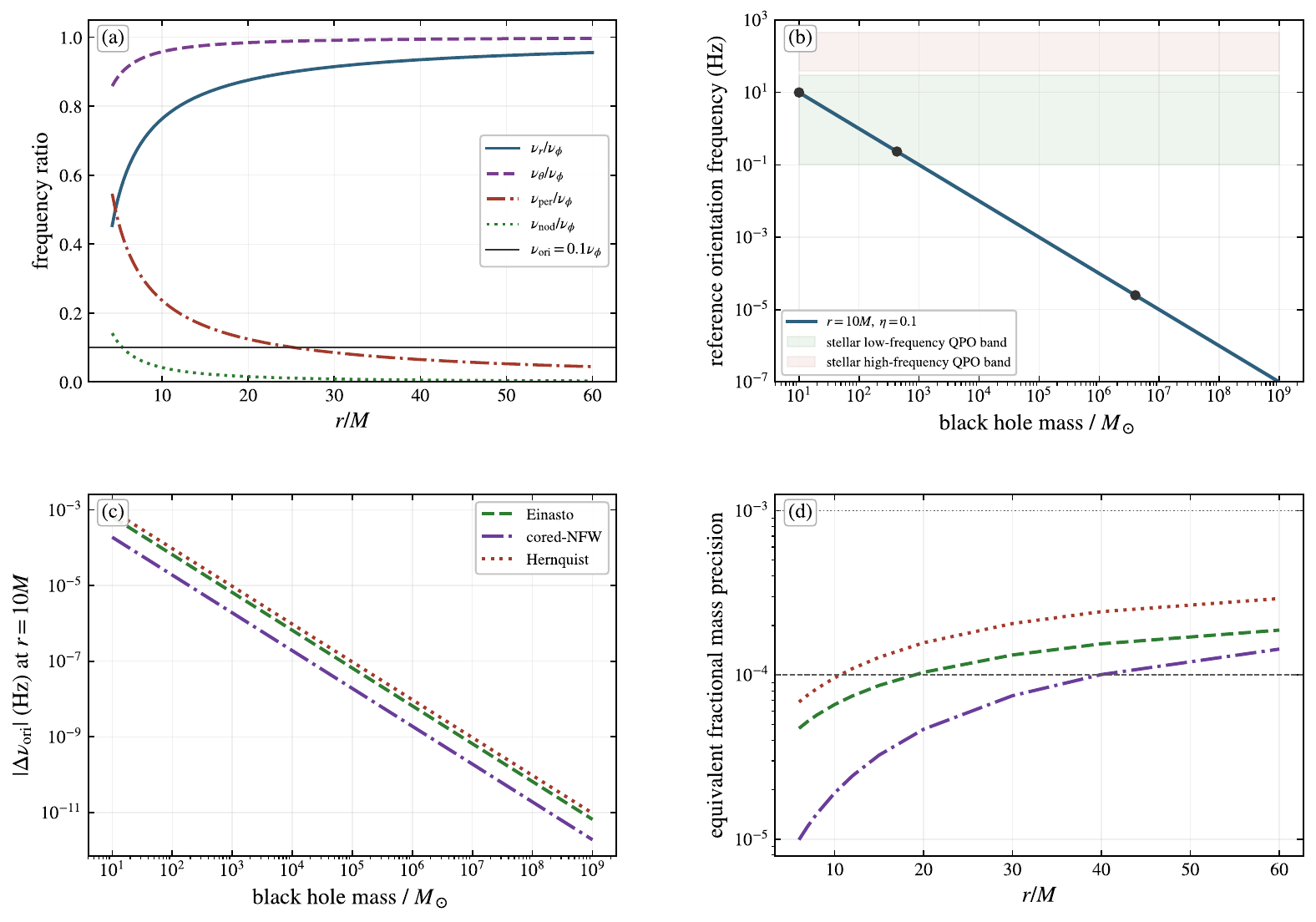}
\caption{Physical timing context. (a) Kerr orbital, epicyclic, periastron, nodal, and illustrative orientation-frequency ratios. (b) Mass scaling for the example $\eta=0.1$ clock at $r=10M$. The shaded stellar QPO bands are only visual guides. (c) Absolute profile displacement for the same closure. (d) Fractional mass precision that would give an equivalent change in one absolute frequency.}
\label{fig:context}
\end{figure}

{The synthetic tests use dimensionless frequency ratios. To compare their scale with standard compact-object timing channels, I now convert the clock to physical units while keeping the unknown body rotation explicit. I set}
\begin{equation}
 \nu_{\rm ori,\Kerr}=\eta\nu_\phi,
\label{eq:eta}
\end{equation}
where $\eta$ is a nuisance parameter. For a prograde Kerr orbit,
\begin{equation}
 \nu_\phi=\frac{c^3}{2\pi GM}
 \frac{1}{(r/M)^{3/2}+\chi}.
\label{eq:nuphi}
\end{equation}
Figure~\ref{fig:context} uses $\eta=0.1$ only as an example.

At $r=10M$ and $\chi=0.8$, a $10M_\odot$ black hole has $\nu_\phi\simeq99.7$ Hz, so the illustrative orientation clock is $9.97$ Hz. The absolute Einasto, cored-NFW, and Hernquist shifts are then $-6.58\times10^{-4}$, $-1.90\times10^{-4}$, and $-9.57\times10^{-4}$ Hz. For $428M_\odot$, the same clock is $0.233$ Hz, below the $3.32\pm0.06$ and $5.07\pm0.06$ Hz peaks reported for M82 X-1 \cite{Pasham:2014mxa}. For $4\times10^6M_\odot$, it is $2.49\times10^{-5}$ Hz. The roughly one-hour QPO in RE J1034+396 gives another useful scale \cite{Gierlinski:2008,Alston:2014}. Because $\eta$, radius, coherence, and emission are unknown, these numbers should be read only as scale comparisons.

The fractional centroid uncertainties of the two M82 X-1 peaks are about $1.8$ and $1.2$ per cent. They are roughly 60 and 40 times larger than the largest shift in the grid. In RE J1034+396, the hard-band QPO changed from $2.47\times10^{-4}$ to $2.83\times10^{-4}$ Hz in ten recent XMM--Newton observations \cite{Xia:2024}. This range is about 14 per cent of the mean and is several hundred times larger than the maximum shift found here. These measurements do not constrain the present model directly. They show that ordinary centroid uncertainty and intrinsic frequency evolution can easily hide a $10^{-4}$ environmental correction.

The DM shifts are equivalent to fractional mass changes of about $10^{-5}$ to $3\times10^{-4}$ in the explored grid. Small changes in radius, spin, or the body parameter $\eta$ can be as large or larger. The orientation clock may also overlap orbital, epicyclic, resonance, diskoseismic, and LT bands \cite{Stella:1999ct,Abramowicz:2001bi,Kato:2008,Ingram:2009eb,Motta:2014qpa}. Assigning an observed QPO to this clock would require all of these channels to be compared in one source-level likelihood.

\section{Discussion and Limitations}
\label{sec:discussion}

\subsection{Interpretation of the paired shift}

{The paired fractional shift is more robust than the absolute flip frequency. The absolute clock depends strongly on the moments of inertia, the initial state, and the quadrupole closure. Pairing each DM run with a Kerr run that uses the same body removes much of this preparation dependence. This is why the environmental response stays smooth even where the intrinsic clock changes rapidly.}

{Equation~\eqref{eq:factorization} explains the remaining structure. In the reduced model, the $u^t$ channel is independent of the body parameters, while the proper-time channel contains the tidal coupling. The $u^t$ channel dominates for the baseline body. The direct $K_{\rm tidal}$ scan shows that a stronger coupling can change this balance.}

\subsection{Scope of the forecast}

{The forecast asks whether the paired shift could be measured if a coherent orientation signal existed. It adds modulation, noise, cadence, coherence, and a prior on the Kerr clock. Published QPO measurements only set the scale. No real source is fitted.}

\subsection{Main limitations}

The rotating DM metric is phenomenological. I have not derived it from a specified stress-energy tensor. A spherical profile may also cease to be self-consistent when rotation is introduced. The profile names label the mass functions used in the ansatz; they do not define a unique strong-field particle model.

The coherent element is a reduced object, not a rigid disk. Pressure, magnetic fields, shear, viscosity, cooling, shocks, turbulence, self-gravity, mass exchange, and disruption are absent. These effects control whether the element can form and keep phase coherence. A GRMHD or local shearing-fluid calculation is needed for that question.

{The tidal torque is not a complete covariant multipole evolution. A full Mathisson--Papapetrou--Dixon model would require a center-of-mass condition, a covariant quadrupole law, and a self-consistent orbit \cite{Dixon:1974c,EhlersRudolph:1977,Bini:2015zya,SteinhoffPuetzfeld:2012}. Here $K_{\rm tidal}$ is only an effective sensitivity. The orbit is also restricted to the circular equatorial case. Eccentricity, inclination, radial drift, or a changing orbit would add new frequencies and could mix the orientation clock with LT and epicyclic motion.}

{The projection maps are local orthographic maps, and the synthetic light curve is a harmonic proxy. They do not include lensing, Doppler boosting, opacity, temperature, synchrotron emission, or radiative transfer. The long phase curves also assume ideal coherence. Dissipation and stochastic forcing may stop the clock well before 1000 cycles, so coherence is likely to be the main observational difficulty.}

Finally, $\mu_{\DM}\le0.05$ is the controlled numerical range used in this study. A realistic halo may be much weaker, while a spike may have a steeper inner slope. The near-linear scaling with $\mu_{\DM}$ is valid only inside the calculated range.

\section{Conclusions}
\label{sec:conclusion}

I have studied how the intrinsic orientation clock of a coherent triaxial element responds to a weakly dark-matter-dressed rotating geometry. The clock comes from the intermediate-axis instability. The surrounding matter does not create the instability; it changes the local curvature and the conversion between proper time and stationary coordinate time.

The exact factorization in Eq.~\eqref{eq:factorization} provides the organizing relation. In the direct grid of 600 dark matter models, the total frequency shift is negative and ranges from about $-3\times10^{-7}$ to $-3\times10^{-4}$. It is nearly linear in the adopted halo normalization. For the fiducial body, the time-conversion channel dominates, although the direct coupling scan shows that a stronger quadrupolar response can increase the proper-time contribution.

The absolute flip frequency remains sensitive to the moments of inertia and the initial state. The paired fractional shift is much more stable, which makes it the cleaner model output. For the chosen profiles and normalization, Hernquist gives the largest shift and cored-NFW the smallest. The time-domain solutions show how these small differences accumulate as a phase drift.

The synthetic timing study sets the observational scale. A shift near $10^{-4}$ can be recovered in a favorable experiment only when the signal is narrow, coherent, well sampled, and compared with an independently constrained Kerr baseline. If the baseline is free, ordinary uncertainties in the body, orbit, mass, spin, and other relativistic timing channels can absorb the effect. {The result is therefore a theory-to-observation framework, not a direct dark matter constraint.} A covariant multipole model, a fluid calculation of the coherence time, and relativistic photon propagation are the main steps needed before source-level inference becomes justified.

\section*{Acknowledgments}
The author acknowledges financial support from Agencia Nacional de Investigación y Desarrollo (ANID) - Chile, through the FONDECYT postdoctoral Grant No.~3260029. 

\section*{Data and code availability}
No proprietary observational data were used. The numerical tables, validation files, figure data, and code needed to reproduce the results are available from the author upon reasonable request.

\bibliographystyle{ws-ijgmmp}
\bibliography{dark_flip_ijgmmp_framework}

@book{Goldstein:2002,
  author    = {Goldstein, Herbert and Poole, Charles P. and Safko, John L.},
  title     = {Classical Mechanics},
  edition   = {3},
  publisher = {Addison-Wesley},
  address   = {San Francisco},
  year      = {2002}
}

@book{Landau:1976,
  author    = {Landau, L. D. and Lifshitz, E. M.},
  title     = {Mechanics},
  edition   = {3},
  publisher = {Butterworth-Heinemann},
  address   = {Oxford},
  year      = {1976}
}

@article{Mathisson:1937zz,
  author  = {Mathisson, Myron},
  title   = {{Neue Mechanik materieller Systeme}},
  journal = {Acta Phys. Polon.},
  volume  = {6},
  pages   = {163--200},
  year    = {1937}
}

@article{Papapetrou:1951pa,
  author  = {Papapetrou, A.},
  title   = {{Spinning test-particles in general relativity. I}},
  journal = {Proc. Roy. Soc. Lond. A},
  volume  = {209},
  pages   = {248--258},
  year    = {1951},
  doi     = {10.1098/rspa.1951.0200}
}

@article{Dixon:1970zza,
  author  = {Dixon, W. G.},
  title   = {{Dynamics of extended bodies in general relativity. I. Momentum and angular momentum}},
  journal = {Proc. Roy. Soc. Lond. A},
  volume  = {314},
  pages   = {499--527},
  year    = {1970},
  doi     = {10.1098/rspa.1970.0020}
}

@article{Dixon:1974c,
  author  = {Dixon, W. G.},
  title   = {{Dynamics of extended bodies in general relativity. III. Equations of motion}},
  journal = {Phil. Trans. Roy. Soc. Lond. A},
  volume  = {277},
  pages   = {59--119},
  year    = {1974},
  doi     = {10.1098/rsta.1974.0046}
}

@article{Kerr:1963ud,
  author  = {Kerr, Roy P.},
  title   = {{Gravitational field of a spinning mass as an example of algebraically special metrics}},
  journal = {Phys. Rev. Lett.},
  volume  = {11},
  pages   = {237--238},
  year    = {1963},
  doi     = {10.1103/PhysRevLett.11.237}
}

@article{Carter:1968rr,
  author  = {Carter, Brandon},
  title   = {{Global structure of the Kerr family of gravitational fields}},
  journal = {Phys. Rev.},
  volume  = {174},
  pages   = {1559--1571},
  year    = {1968},
  doi     = {10.1103/PhysRev.174.1559}
}

@article{Bardeen:1972fi,
  author  = {Bardeen, James M. and Press, William H. and Teukolsky, Saul A.},
  title   = {{Rotating black holes: locally nonrotating frames, energy extraction, and scalar synchrotron radiation}},
  journal = {Astrophys. J.},
  volume  = {178},
  pages   = {347},
  year    = {1972},
  doi     = {10.1086/151796}
}

@book{Chandrasekhar:1983,
  author    = {Chandrasekhar, S.},
  title     = {{The Mathematical Theory of Black Holes}},
  publisher = {Oxford University Press},
  address   = {Oxford},
  year      = {1983}
}

@article{Newman:1965tw,
  author  = {Newman, E. T. and Janis, A. I.},
  title   = {{Note on the Kerr spinning-particle metric}},
  journal = {J. Math. Phys.},
  volume  = {6},
  pages   = {915--917},
  year    = {1965},
  doi     = {10.1063/1.1704350}
}

@article{AzregAinou:2014pra,
  author        = {Azreg-A\"inou, Mustapha},
  title         = {{Generating rotating regular black hole solutions without complexification}},
  journal       = {Phys. Rev. D},
  volume        = {90},
  number        = {6},
  pages         = {064041},
  year          = {2014},
  doi           = {10.1103/PhysRevD.90.064041},
  eprint        = {1405.2569},
  archivePrefix = {arXiv},
  primaryClass  = {gr-qc}
}

@article{Bini:2015zya,
  author        = {Bini, Donato and Faye, Guillaume and Geralico, Andrea},
  title         = {{Dynamics of extended bodies in a Kerr spacetime with spin-induced quadrupole tensor}},
  journal       = {Phys. Rev. D},
  volume        = {92},
  number        = {10},
  pages         = {104003},
  year          = {2015},
  doi           = {10.1103/PhysRevD.92.104003},
  eprint        = {1507.07441},
  archivePrefix = {arXiv},
  primaryClass  = {gr-qc}
}

@article{Han:2016cdm,
  author        = {Han, Wen-Biao and Cheng, Ran},
  title         = {{Dynamics of extended bodies with spin-induced quadrupole in Kerr spacetime: generic orbits}},
  journal       = {Gen. Rel. Grav.},
  volume        = {49},
  pages         = {48},
  year          = {2017},
  doi           = {10.1007/s10714-017-2217-1},
  eprint        = {1611.07602},
  archivePrefix = {arXiv},
  primaryClass  = {gr-qc}
}

@article{Einasto:1965,
  author  = {Einasto, Jaan},
  title   = {{On the construction of a composite model for the galaxy and on the determination of the system of galactic parameters}},
  journal = {Trudy Astrofizicheskogo Instituta Alma-Ata},
  volume  = {5},
  pages   = {87--100},
  year    = {1965}
}

@article{Navarro:1996gj,
  author        = {Navarro, Julio F. and Frenk, Carlos S. and White, Simon D. M.},
  title         = {{The structure of cold dark matter halos}},
  journal       = {Astrophys. J.},
  volume        = {462},
  pages         = {563--575},
  year          = {1996},
  doi           = {10.1086/177173},
  eprint        = {astro-ph/9508025},
  archivePrefix = {arXiv}
}

@article{Navarro:1996he,
  author        = {Navarro, Julio F. and Frenk, Carlos S. and White, Simon D. M.},
  title         = {{A universal density profile from hierarchical clustering}},
  journal       = {Astrophys. J.},
  volume        = {490},
  pages         = {493--508},
  year          = {1997},
  doi           = {10.1086/304888},
  eprint        = {astro-ph/9611107},
  archivePrefix = {arXiv}
}

@article{Hernquist:1990be,
  author  = {Hernquist, Lars},
  title   = {{An analytical model for spherical galaxies and bulges}},
  journal = {Astrophys. J.},
  volume  = {356},
  pages   = {359--364},
  year    = {1990},
  doi     = {10.1086/168845}
}

@article{Burkert:1995yz,
  author        = {Burkert, A.},
  title         = {{The structure of dark matter halos in dwarf galaxies}},
  journal       = {Astrophys. J. Lett.},
  volume        = {447},
  pages         = {L25--L28},
  year          = {1995},
  doi           = {10.1086/309560},
  eprint        = {astro-ph/9504041},
  archivePrefix = {arXiv}
}

@article{Gondolo:1999ef,
  author        = {Gondolo, Paolo and Silk, Joseph},
  title         = {{Dark matter annihilation at the galactic center}},
  journal       = {Phys. Rev. Lett.},
  volume        = {83},
  pages         = {1719--1722},
  year          = {1999},
  doi           = {10.1103/PhysRevLett.83.1719},
  eprint        = {astro-ph/9906391},
  archivePrefix = {arXiv}
}

@article{Ullio:2001fb,
  author        = {Ullio, Piero and Zhao, HongSheng and Kamionkowski, Marc},
  title         = {{Dark-matter spike at the galactic center?}},
  journal       = {Phys. Rev. D},
  volume        = {64},
  pages         = {043504},
  year          = {2001},
  doi           = {10.1103/PhysRevD.64.043504},
  eprint        = {astro-ph/0101481},
  archivePrefix = {arXiv}
}

@article{Sadeghian:2013laa,
  author        = {Sadeghian, Laleh and Ferrer, Francesc and Will, Clifford M.},
  title         = {{Dark matter distributions around massive black holes: a general relativistic analysis}},
  journal       = {Phys. Rev. D},
  volume        = {88},
  number        = {6},
  pages         = {063522},
  year          = {2013},
  doi           = {10.1103/PhysRevD.88.063522},
  eprint        = {1305.2619},
  archivePrefix = {arXiv},
  primaryClass  = {astro-ph.GA}
}

@article{Cardoso:2021wlq,
  author        = {Cardoso, Vitor and Destounis, Kyriakos and Duque, Francisco and Macedo, Rodrigo Panosso and Maselli, Andrea},
  title         = {{Black holes in galaxies: environmental impact on gravitational-wave generation and propagation}},
  journal       = {Phys. Rev. D},
  volume        = {105},
  number        = {6},
  pages         = {L061501},
  year          = {2022},
  doi           = {10.1103/PhysRevD.105.L061501},
  eprint        = {2109.00005},
  archivePrefix = {arXiv},
  primaryClass  = {gr-qc}
}

@article{Figueiredo:2023gas,
  author        = {Figueiredo, Enzo and Maselli, Andrea and Cardoso, Vitor},
  title         = {{Black holes surrounded by generic dark matter profiles: appearance and gravitational-wave emission}},
  journal       = {Phys. Rev. D},
  volume        = {107},
  number        = {10},
  pages         = {104033},
  year          = {2023},
  doi           = {10.1103/PhysRevD.107.104033},
  eprint        = {2303.08183},
  archivePrefix = {arXiv},
  primaryClass  = {gr-qc}
}

@article{Kavanagh:2020cfn,
  author        = {Kavanagh, Bradley J. and Nichols, David A. and Bertone, Gianfranco and Gaggero, Daniele},
  title         = {{Detecting dark matter around black holes with gravitational waves: effects of dark-matter dynamics on the gravitational waveform}},
  journal       = {Phys. Rev. D},
  volume        = {102},
  number        = {8},
  pages         = {083006},
  year          = {2020},
  doi           = {10.1103/PhysRevD.102.083006},
  eprint        = {2002.12811},
  archivePrefix = {arXiv},
  primaryClass  = {gr-qc}
}

@article{Barausse:2014tra,
  author        = {Barausse, Enrico and Cardoso, Vitor and Pani, Paolo},
  title         = {{Environmental effects for gravitational-wave astrophysics}},
  journal       = {Phys. Rev. D},
  volume        = {89},
  number        = {10},
  pages         = {104059},
  year          = {2014},
  doi           = {10.1103/PhysRevD.89.104059},
  eprint        = {1404.7149},
  archivePrefix = {arXiv},
  primaryClass  = {gr-qc}
}

@article{Xu:2018wow,
  author        = {Xu, Zhaoyi and Hou, Xian and Gong, Xiaobo and Wang, Jiancheng},
  title         = {{Black hole space-time in dark matter halo}},
  journal       = {JCAP},
  volume        = {09},
  pages         = {038},
  year          = {2018},
  doi           = {10.1088/1475-7516/2018/09/038},
  eprint        = {1803.00767},
  archivePrefix = {arXiv},
  primaryClass  = {gr-qc}
}

@article{Xu:2020bxi,
  author        = {Xu, Zhaoyi and Gong, Xiaobo and Zhang, Shuang-Nan},
  title         = {{Black hole immersed dark matter halo}},
  journal       = {Phys. Rev. D},
  volume        = {101},
  number        = {2},
  pages         = {024029},
  year          = {2020},
  doi           = {10.1103/PhysRevD.101.024029},
  eprint        = {2002.00730},
  archivePrefix = {arXiv},
  primaryClass  = {gr-qc}
}

@article{Jusufi:2019ltj,
  author        = {Jusufi, Kimet and Jamil, Mubasher and Salucci, Paolo and Zhu, Tao and Haroon, Sumarna},
  title         = {{Black hole surrounded by a dark matter halo in the M87 galactic center and its identification with shadow images}},
  journal       = {Phys. Rev. D},
  volume        = {100},
  number        = {4},
  pages         = {044012},
  year          = {2019},
  doi           = {10.1103/PhysRevD.100.044012},
  eprint        = {1905.11803},
  archivePrefix = {arXiv},
  primaryClass  = {physics.gen-ph}
}

@article{Stella:1999ct,
  author        = {Stella, Luigi and Vietri, Mario},
  title         = {{Lense-Thirring precession and quasi-periodic oscillations in low-mass X-ray binaries}},
  journal       = {Phys. Rev. Lett.},
  volume        = {82},
  pages         = {17--20},
  year          = {1999},
  doi           = {10.1103/PhysRevLett.82.17},
  eprint        = {astro-ph/9812124},
  archivePrefix = {arXiv}
}

@article{Remillard:2006fc,
  author        = {Remillard, Ronald A. and McClintock, Jeffrey E.},
  title         = {{X-ray properties of black-hole binaries}},
  journal       = {Ann. Rev. Astron. Astrophys.},
  volume        = {44},
  pages         = {49--92},
  year          = {2006},
  doi           = {10.1146/annurev.astro.44.051905.092532},
  eprint        = {astro-ph/0606352},
  archivePrefix = {arXiv}
}

@article{Ingram:2009eb,
  author        = {Ingram, Adam and Done, Chris and Fragile, P. Chris},
  title         = {{Low-frequency quasi-periodic oscillations spectra and Lense-Thirring precession}},
  journal       = {Mon. Not. Roy. Astron. Soc.},
  volume        = {397},
  pages         = {L101--L105},
  year          = {2009},
  doi           = {10.1111/j.1745-3933.2009.00693.x},
  eprint        = {0901.1238},
  archivePrefix = {arXiv},
  primaryClass  = {astro-ph.HE}
}

@article{Motta:2014qpa,
  author        = {Motta, S. E. and Belloni, T. M. and Stella, L. and Mu\~noz-Darias, T. and Fender, R.},
  title         = {{Precise mass and spin measurements for a stellar-mass black hole through X-ray timing: the case of GRO J1655--40}},
  journal       = {Mon. Not. Roy. Astron. Soc.},
  volume        = {437},
  pages         = {2554--2565},
  year          = {2014},
  doi           = {10.1093/mnras/stt2068},
  eprint        = {1309.3652},
  archivePrefix = {arXiv},
  primaryClass  = {astro-ph.HE}
}

@article{Pasham:2014mxa,
  author  = {Pasham, Dheeraj R. and Strohmayer, Tod E. and Mushotzky, Richard F.},
  title   = {{A 400-solar-mass black hole in the galaxy M82}},
  journal = {Nature},
  volume  = {513},
  pages   = {74--76},
  year    = {2014},
  doi     = {10.1038/nature13710}
}

@incollection{vanDerKlis:2006,
  author    = {van der Klis, Michiel},
  title     = {{Rapid X-ray variability}},
  booktitle = {{Compact Stellar X-ray Sources}},
  editor    = {Lewin, Walter H. G. and van der Klis, Michiel},
  publisher = {Cambridge University Press},
  pages     = {39--112},
  year      = {2006},
  eprint    = {astro-ph/0410551},
  archivePrefix = {arXiv}
}

@article{Abramowicz:2001bi,
  author        = {Abramowicz, Marek A. and Kluzniak, Wlodek},
  title         = {{A precise determination of black hole spin in GRO J1655-40}},
  journal       = {Astron. Astrophys.},
  volume        = {374},
  pages         = {L19--L20},
  year          = {2001},
  doi           = {10.1051/0004-6361:20010791},
  eprint        = {astro-ph/0105077},
  archivePrefix = {arXiv}
}

@book{Kato:2008,
  author    = {Kato, Shoji and Fukue, Jun and Mineshige, Shin},
  title     = {{Black-Hole Accretion Disks: Towards a New Paradigm}},
  publisher = {Kyoto University Press},
  address   = {Kyoto},
  year      = {2008}
}

@article{Nowak:1999,
  author        = {Nowak, Michael A. and Wilms, Joern and Dove, James B.},
  title         = {{Low-frequency quasi-periodic oscillations in black hole candidates}},
  journal       = {Astrophys. J.},
  volume        = {517},
  pages         = {355--366},
  year          = {1999},
  doi           = {10.1086/307184},
  eprint        = {astro-ph/9810396},
  archivePrefix = {arXiv}
}

@article{Virtanen:2019joe,
  author  = {Virtanen, Pauli and others},
  title   = {{SciPy 1.0: fundamental algorithms for scientific computing in Python}},
  journal = {Nature Meth.},
  volume  = {17},
  pages   = {261--272},
  year    = {2020},
  doi     = {10.1038/s41592-019-0686-2}
}

@article{Harris:2020xlr,
  author  = {Harris, Charles R. and others},
  title   = {{Array programming with NumPy}},
  journal = {Nature},
  volume  = {585},
  pages   = {357--362},
  year    = {2020},
  doi     = {10.1038/s41586-020-2649-2}
}

@article{Hunter:2007,
  author  = {Hunter, John D.},
  title   = {{Matplotlib: A 2D graphics environment}},
  journal = {Comput. Sci. Eng.},
  volume  = {9},
  pages   = {90--95},
  year    = {2007},
  doi     = {10.1109/MCSE.2007.55}
}

@article{VanDamme:2017,
  author = {Van Damme, L. and Marde\v{s}i\'c, P. and Sugny, D.},
  title = {The tennis racket effect in a three-dimensional rigid body},
  journal = {Physica D},
  volume = {338},
  pages = {17--25},
  year = {2017},
  doi = {10.1016/j.physd.2016.07.010},
  eprint = {1606.08237},
  archivePrefix = {arXiv},
  primaryClass = {physics.class-ph}
}

@article{Mardesic:2020,
  author = {Marde\v{s}i\'c, P. and Van Damme, L. and Guti\'errez Guill\'en, G. J. and Sugny, D.},
  title = {Geometric origin of the tennis racket effect},
  journal = {Phys. Rev. Lett.},
  volume = {125},
  pages = {064301},
  year = {2020},
  doi = {10.1103/PhysRevLett.125.064301},
  eprint = {2003.13539},
  archivePrefix = {arXiv},
  primaryClass = {physics.class-ph}
}

@article{deLaTorre:2024,
  author = {de la Torre, J. A. and Espa\~nol, P.},
  title = {Internal dissipation in the {Dzhanibekov} effect},
  journal = {Eur. J. Mech. A Solids},
  volume = {106},
  pages = {105298},
  year = {2024},
  doi = {10.1016/j.euromechsol.2024.105298},
  eprint = {2312.15448},
  archivePrefix = {arXiv},
  primaryClass = {physics.class-ph}
}

@article{Bini:2009,
  author = {Bini, D. and Fortini, P. and Geralico, A. and Ortolan, A.},
  title = {Quadrupole effects on the motion of extended bodies in {Kerr} spacetime},
  journal = {Class. Quantum Grav.},
  volume = {27},
  pages = {185003},
  year = {2010},
  doi = {10.1088/0264-9381/27/18/185003},
  eprint = {0910.2842},
  archivePrefix = {arXiv},
  primaryClass = {gr-qc}
}

@article{BiniGeralico:2014,
  author = {Bini, D. and Geralico, A.},
  title = {Extended bodies in a {Kerr} spacetime: exploring the role of a general quadrupole tensor},
  journal = {Phys. Rev. D},
  volume = {89},
  pages = {044013},
  year = {2014},
  doi = {10.1103/PhysRevD.89.044013},
  eprint = {1408.5484},
  archivePrefix = {arXiv},
  primaryClass = {gr-qc}
}

@article{IngramMotta:2020,
  author = {Ingram, A. R. and Motta, S. E.},
  title = {A review of quasi-periodic oscillations from black hole {X}-ray binaries: observation and theory},
  journal = {New Astronomy Reviews},
  volume = {85},
  pages = {101524},
  year = {2020},
  doi = {10.1016/j.newar.2020.101524},
  eprint = {2001.08758},
  archivePrefix = {arXiv},
  primaryClass = {astro-ph.HE}
}

@article{Gierlinski:2008,
  author = {Gierli\'nski, M. and Middleton, M. and Ward, M. and Done, C.},
  title = {A one-hour {X}-ray periodicity in an active galaxy},
  journal = {Nature},
  volume = {455},
  pages = {369--371},
  year = {2008},
  doi = {10.1038/nature07277},
  eprint = {0807.1899},
  archivePrefix = {arXiv},
  primaryClass = {astro-ph}
}

@article{Alston:2014,
  author = {Alston, W. N. and Markeviciute, J. and Kara, E. and Fabian, A. C. and Middleton, M.},
  title = {Detection of a quasi-periodic oscillation in five {XMM--Newton} observations of {RE J1034+396}},
  journal = {Mon. Not. R. Astron. Soc.},
  volume = {445},
  pages = {L16--L20},
  year = {2014},
  doi = {10.1093/mnrasl/slu127},
  eprint = {1407.7657},
  archivePrefix = {arXiv},
  primaryClass = {astro-ph.HE}
}

@article{Lomb:1976,
  author = {Lomb, N. R.},
  title = {Least-squares frequency analysis of unequally spaced data},
  journal = {Astrophys. Space Sci.},
  volume = {39},
  pages = {447--462},
  year = {1976},
  doi = {10.1007/BF00648343}
}

@article{Scargle:1982,
  author = {Scargle, J. D.},
  title = {Studies in astronomical time series analysis. {II}. {Statistical} aspects of spectral analysis of unevenly spaced data},
  journal = {Astrophys. J.},
  volume = {263},
  pages = {835--853},
  year = {1982},
  doi = {10.1086/160554}
}

@article{Vaughan:2005,
  author = {Vaughan, S.},
  title = {A simple test for periodic signals in red noise},
  journal = {Astron. Astrophys.},
  volume = {431},
  pages = {391--403},
  year = {2005},
  doi = {10.1051/0004-6361:20041453},
  eprint = {astro-ph/0412697},
  archivePrefix = {arXiv}
}

@article{Schwarz:1978,
  author = {Schwarz, G.},
  title = {Estimating the dimension of a model},
  journal = {Annals of Statistics},
  volume = {6},
  pages = {461--464},
  year = {1978},
  doi = {10.1214/aos/1176344136}
}

@article{Xia:2024,
  author        = {Xia, Ruisong and Liu, Hao and Xue, Yongquan},
  title         = {{First observational evidence for an interconnected evolution between time lag and QPO frequency among AGNs}},
  journal       = {Astrophys. J. Lett.},
  volume        = {961},
  number        = {2},
  pages         = {L32},
  year          = {2024},
  doi           = {10.3847/2041-8213/ad1bf2},
  eprint        = {2401.04926},
  archivePrefix = {arXiv},
  primaryClass  = {astro-ph.HE}
}

@article{ManasseMisner:1963,
  author  = {Manasse, F. K. and Misner, C. W.},
  title   = {Fermi normal coordinates and some basic concepts in differential geometry},
  journal = {J. Math. Phys.},
  volume  = {4},
  number  = {6},
  pages   = {735--745},
  year    = {1963},
  doi     = {10.1063/1.1724316}
}

@article{Marck:1983,
  author  = {Marck, Jean-Alain},
  title   = {Solution to the equations of parallel transport in {Kerr} geometry; tidal tensor},
  journal = {Proc. R. Soc. Lond. A},
  volume  = {385},
  number  = {1789},
  pages   = {431--438},
  year    = {1983},
  doi     = {10.1098/rspa.1983.0021}
}

@article{EhlersRudolph:1977,
  author  = {Ehlers, J. and Rudolph, E.},
  title   = {Dynamics of extended bodies in general relativity: center-of-mass description and quasirigidity},
  journal = {Gen. Relativ. Gravit.},
  volume  = {8},
  pages   = {197--217},
  year    = {1977},
  doi     = {10.1007/BF00763547}
}

@article{SteinhoffPuetzfeld:2012,
  author        = {Steinhoff, Jan and Puetzfeld, Dirk},
  title         = {Influence of internal structure on the motion of test bodies in extreme mass ratio situations},
  journal       = {Phys. Rev. D},
  volume        = {86},
  pages         = {044033},
  year          = {2012},
  doi           = {10.1103/PhysRevD.86.044033},
  eprint        = {1205.3926},
  archivePrefix = {arXiv},
  primaryClass  = {gr-qc}
}

@article{Montgomery:1991,
  author  = {Montgomery, Richard},
  title   = {How much does the rigid body rotate? {A} {Berry}'s phase from the 18th century},
  journal = {Am. J. Phys.},
  volume  = {59},
  number  = {5},
  pages   = {394--398},
  year    = {1991},
  doi     = {10.1119/1.16514}
}

@incollection{BachettiHuppenkothen:2024,
  author    = {Bachetti, Matteo and Huppenkothen, Daniela},
  title     = {Fourier methods},
  booktitle = {Handbook of X-ray and Gamma-ray Astrophysics},
  editor    = {Bambi, Cosimo and Santangelo, Andrea},
  publisher = {Springer},
  address   = {Singapore},
  year      = {2024},
  doi       = {10.1007/978-981-19-6960-7_137},
  eprint    = {2209.07954},
  archivePrefix = {arXiv},
  primaryClass  = {astro-ph.IM}
}

@article{VanderPlas:2018,
  author        = {VanderPlas, Jacob T.},
  title         = {Understanding the {Lomb--Scargle} periodogram},
  journal       = {Astrophys. J. Suppl. Ser.},
  volume        = {236},
  number        = {1},
  pages         = {16},
  year          = {2018},
  doi           = {10.3847/1538-4365/aab766},
  eprint        = {1703.09824},
  archivePrefix = {arXiv},
  primaryClass  = {astro-ph.IM}
}

@article{FritschCarlson:1980,
  author  = {Fritsch, F. N. and Carlson, R. E.},
  title   = {Monotone piecewise cubic interpolation},
  journal = {SIAM J. Numer. Anal.},
  volume  = {17},
  number  = {2},
  pages   = {238--246},
  year    = {1980},
  doi     = {10.1137/0717021}
}

@book{Hairer:1993,
  author    = {Hairer, Ernst and Norsett, Syvert P. and Wanner, Gerhard},
  title     = {Solving Ordinary Differential Equations I: Nonstiff Problems},
  edition   = {2},
  publisher = {Springer},
  address   = {Berlin},
  year      = {1993},
  doi       = {10.1007/978-3-540-78862-1}
}

@article{Welch:1967,
  author  = {Welch, Peter D.},
  title   = {The use of fast {Fourier} transform for the estimation of power spectra: A method based on time averaging over short, modified periodograms},
  journal = {IEEE Trans. Audio Electroacoust.},
  volume  = {15},
  number  = {2},
  pages   = {70--73},
  year    = {1967},
  doi     = {10.1109/TAU.1967.1161901}
}

@article{Gammie:2001,
  author = {Gammie, Charles F.},
  title = {Nonlinear Outcome of Gravitational Instability in Cooling, Gaseous Disks},
  journal = {Astrophys. J.},
  volume = {553},
  pages = {174--183},
  year = {2001},
  doi = {10.1086/320631},
  eprint = {astro-ph/0101501},
  archivePrefix = {arXiv}
}

@article{Cossins:2009,
  author = {Cossins, Peter and Lodato, Giuseppe and Clarke, C. J.},
  title = {Characterizing the gravitational instability in cooling accretion discs},
  journal = {Mon. Not. Roy. Astron. Soc.},
  volume = {393},
  number = {4},
  pages = {1157--1173},
  year = {2009},
  doi = {10.1111/j.1365-2966.2008.14275.x},
  eprint = {0811.3629},
  archivePrefix = {arXiv},
  primaryClass = {astro-ph}
}

@article{Schnittman:2005,
  author = {Schnittman, Jeremy D.},
  title = {Interpreting the High-Frequency Quasi-periodic Oscillation Power Spectra of Accreting Black Holes},
  journal = {Astrophys. J.},
  volume = {621},
  pages = {940--950},
  year = {2005},
  doi = {10.1086/427149},
  eprint = {astro-ph/0407179},
  archivePrefix = {arXiv}
}

@article{Ripperda:2020,
  author = {Ripperda, Bart and Bacchini, Fabio and Philippov, Alexander A.},
  title = {Magnetic Reconnection and Hot Spot Formation in Black Hole Accretion Disks},
  journal = {Astrophys. J.},
  volume = {900},
  number = {2},
  pages = {100},
  year = {2020},
  doi = {10.3847/1538-4357/ababab},
  eprint = {2003.04330},
  archivePrefix = {arXiv},
  primaryClass = {astro-ph.HE}
}

\end{document}